\documentclass[11pt, a4paper]{article}
\usepackage[utf8]{inputenc}
\usepackage[T1]{fontenc}
\usepackage{authblk}
\usepackage[a4paper,top=2.5cm,bottom=2.5cm,left=2.5cm,right=2.5cm,marginparwidth=1.75cm]{geometry}

\usepackage{algorithm}
\usepackage[noend]{algpseudocode}
\usepackage{amsmath, amssymb}
\usepackage{bm} 
\usepackage[numbers]{natbib}
\usepackage{setspace, array}
\usepackage{mathtools}
\usepackage{hyperref}

\newcommand{\vect}[1]{\bm{#1}}
\newcommand{\matr}[1]{\bm{#1}}

\usepackage{multirow}
\newcommand{\yin}[0]{\vect{u}_{in}}
\newcommand{\yintilde}[0]{\tilde{\vect{u}}_{in}}
\DeclareMathOperator*{\argmin}{arg\,min} 
\DeclareMathOperator*{\argmax}{arg\,max}

\date{}

\author[1]{D. E. Ozan}
\author[1]{A. N\'ovoa}
\author[1]{G. Rigas}
\author[1-3,*]{L. Magri}

\title{Data-assimilated model-informed reinforcement learning}

\affil[1]{Department of Aeronautics, Imperial College London, London SW7 2AZ, UK.}
\affil[2]{The Alan Turing Institute, London NW1 2DB, UK.}
\affil[3]{DIMEAS, Politecnico di Torino, Torino 24 10129, Italy.}
\affil[*]{\textit{Corresponding author:} \texttt{l.magri@imperial.ac.uk}}

\usepackage[dvipsnames]{xcolor}
\newcommand{\revise}[1]{{\color{black}{#1}}}

\begin{document}
\maketitle 
\begin{abstract}
The control of \revise{spatio-temporal} chaos is challenging because of high dimensionality and unpredictability. 
Model-free reinforcement learning (RL) discovers optimal control policies by interacting with the system, typically requiring observations of the full physical state.
In practice, sensors often provide only partial and noisy measurements (observations) of the system. 
The objective of this paper is to develop a framework that enables the control of chaotic systems with partial and noisy observability. The proposed method, data-assimilated model-informed reinforcement learning (DA-MIRL), integrates 
(i) low-order models to approximate high-dimensional dynamics;  
(ii) sequential data assimilation to correct the model prediction when observations become available; and 
(iii) an off-policy actor-critic RL algorithm to adaptively learn an optimal control strategy based on the corrected state estimates. 
We test DA-MIRL on the \revise{spatio-temporally} chaotic solutions of the Kuramoto-Sivashinsky equation. We estimate the full state of the environment with 
(i) a physics-based model, here, a coarse-grained model; and 
(ii) a data-driven model, here, the control-aware echo state network, which is proposed in this paper. 
We show that DA-MIRL successfully estimates and suppresses the chaotic dynamics of the environment in real time from partial observations and approximate models. 
This work opens opportunities for the control of partially observable chaotic systems.

\end{abstract}

\section{Introduction}

Chaos emerges in a variety of nonlinear dynamical systems across science and engineering, from weather forecasting to cardiovascular physiology~\citep{ott2008ChaosDynamicalSystems}. 
Characterized by nonlinearity, aperiodic behaviour, and extreme sensitivity to small perturbations, chaotic systems are inherently challenging to predict and control~\citep{ott2008ChaosDynamicalSystems}. 
When the nonlinear interactions propagate disturbances across both spatial and temporal domains, the dynamics become spatio-temporally chaotic, e.g., in turbulent flows~\citep{holmes2012TurbulenceCoherentStructures}. 
Controlling spatio-temporal chaos remains a longstanding challenge with significant engineering implications---e.g., effective control strategies can reduce drag on vehicles, increase wind turbine efficiency, or enhance mixing in combustion processes~\citep{brunton2015ClosedLoopTurbulenceControl}. 
However, predicting these systems involves high-dimensional computationally intensive models, which are not suitable in real-time control applications. 
%
Control is, at its core, an optimization problem, in which we seek a strategy to improve the performance of a system (e.g., minimizing energy consumption)~\citep{bertsekas2007DynamicProgrammingOptimal}. 
Practically, closed-loop control dynamically adjusts the control inputs (or actuations) based on feedback from sensor measurements~\citep{astrom2008FeedbackSystemsIntroduction}. 
On the one hand, model-driven control relies on a mathematical model that captures the governing physics of the environment to provide the control strategy. 
Optimal control theory offers a principled way to derive
such policies~\citep{bertsekas2007DynamicProgrammingOptimal}, 
and adjoint methods enable gradient-based optimization in high-dimensional systems by efficiently computing sensitivities of the objective functional with respect to all control parameters~\citep{gunzburger2002PerspectivesFlowControl,bewley2001DNSbasedPredictiveControl}. 
Nonetheless, the implementation of optimal control frameworks typically relies on access to measurements of the full state~\citep{luenberger1971IntroductionObservers, kalman1960NewApproachLinear}. 
Full observability is, however, rarely attainable and only a subset of the full state can be measured---especially for chaotic and turbulent fluid systems. In this paper, given partial noisy sensor measurements and a numerical model, we aim to estimate (i.e., approximate or reconstruct) the full state of the system, which, in turn, enables the feedback control~\citep{luenberger1971IntroductionObservers, kalman1960NewApproachLinear}. 

Data assimilation \revise{(DA)} methods solve the inverse problem of inferring the most likely system state by optimally combining model forecasts and observations, whilst accounting for their uncertainties~\citep{evensen2009DataAssimilationEnsemble}. 
Specifically, sequential DA assimilates the observations as they become available, bypassing the need to store and post-process the data, which makes it suitable for real-time state estimation and control. 
%
State estimation via sequential DA is the backbone of observer-based control methods such as Linear-Quadratic-Gaussian (LQG), which combines the Linear-Quadratic Regulator for optimal state-feedback control with the Kalman filter for minimum-variance state estimation in linear time-invariant (LTI) systems. 
However, LQG in general is not adequate for controlling spatio-temporal chaos as the linear and Gaussian assumptions may break down in high-dimensional chaotic regimes.
Further, adjoint-based optimization used in model-driven control is unstable for chaotic systems due the exponential divergence of nearby trajectories~\citep{lea2000SensitivityAnalysisClimate}. 
%

On the other hand, data-driven control---particularly Reinforcement learning (RL)---provides a framework to find optimal control policies directly from experimental data or simulations~\citep{sutton2018ReinforcementLearningIntroduction} and, therefore, has gained traction in flow control problems~\citep[e.g.,][]{rabault2019ArtificialNeuralNetworks,guastoni2023DeepReinforcementLearning,vignon2023RecentAdvancesApplying}. In RL, an agent learns an optimal policy by interacting with an environment to maximize cumulative rewards~\citep{sutton2018ReinforcementLearningIntroduction}. 
Within RL, actor-critic methods are suited for continuous state and action spaces, as they iteratively update both the value function (i.e., the objective function to optimize) and the control policy~\citep{sutton2018ReinforcementLearningIntroduction}.
Traditionally, RL assumes Markovian dynamics and full state observability, i.e., it is a Markov Decision Process (MDP)~\citep{sutton2018ReinforcementLearningIntroduction}. 
When only a subset of the state can be measured, the problem becomes \revise{a} Partially Observed Markov Decision Process (POMDP)~\citep{kaelbling1998PlanningActingPartially}. 
\begin{figure}
  \centering
  \includegraphics[width=\linewidth]{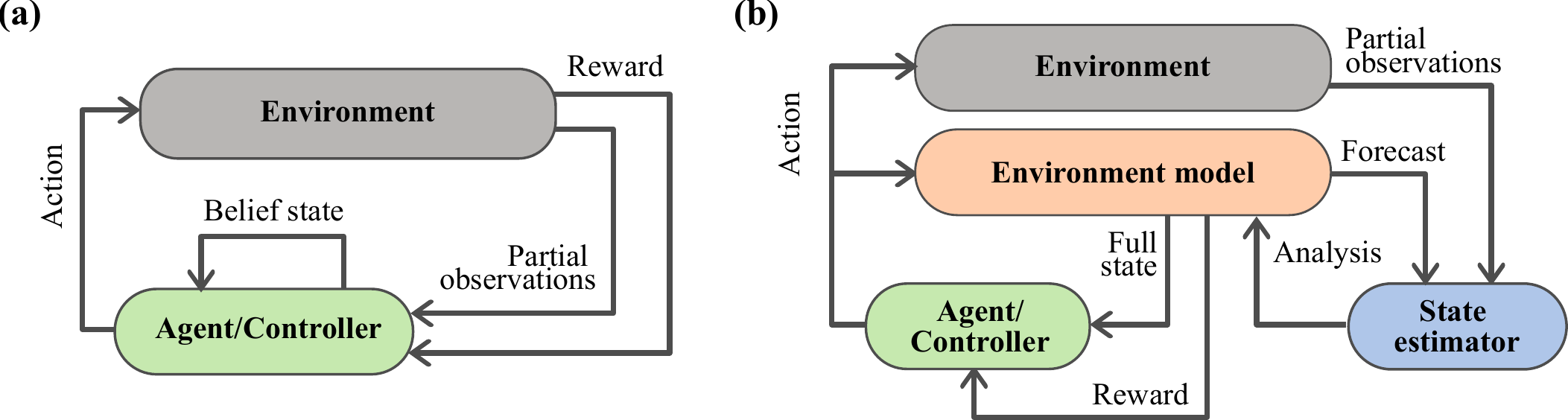}
  \vspace*{-4pt}
  \caption{Three-block schematic of reinforcement learning methods under partial observability. Comparison between (a) model-free RL with belief state and (b) the proposed data-assimilated model-informed RL. }
  \label{fig:methods_compare}
\end{figure}

Using traditional RL agents in POMDP may result in unstable or suboptimal policies~\citep{kaelbling1998PlanningActingPartially}. 
Partial observability in RL has been addressed by  introducing memory into the agent by, for instance, 
employing recurrent neural networks \revise{(RNNs)}, such as Long-Short Term Memory Networks~\citep{verma2018EfficientCollectiveSwimming}, or 
using attention mechanisms~\citep{weissenbacher2025ReinforcementLearningChaotic}. 
Alternatively, POMDPs can be transformed into MDPs by using storing the memory of previous actions and observations in an information vector, which can be fed to the controller as a full state vector~\citep{bertsekas2007DynamicProgrammingOptimal,bucci2019ControlChaoticSystems}. \revise{In this direction, \citet{xia2024ActiveFlowControl} achieve drag reduction of the two-dimensional bluff body from pressure sensor measurements only on the body by the augmentation of the observation history.  In a similar setup, \citet{wang2024DynamicFeaturebasedDeep} learn dynamic features that lift the sparse sensor measurements into a higher dimensional space.}
\revise{These data-stacking approaches} counteract the non-Markovian memory effects introduced by partial observations, which is formalized by the Mori-Zwanzig framework~\citep{gouasmi2017PrioriEstimationMemory}. For chaotic dynamical systems, the use of observation histories is justified by the Takens's embedding theorem---a chaotic attractor can be reconstructed from delayed embeddings of a single observable~\citep{takens1981DetectingStrangeAttractorsa, weissenbacher2025ReinforcementLearningChaotic}. 
However, the dimensionality of the information vector increases with each new observation. This motivates the search for sufficient lower-dimensional statistics, i.e., the belief state~\citep{bertsekas2007DynamicProgrammingOptimal}, which is a probability distribution over all possible system states, given the history of observations and actions. The belief state provides sufficient statistics to encapsulate the information from the observation and action history needed for optimal decision-making under partial observability~\citep{kaelbling1998PlanningActingPartially}. 
The belief state is recursively updated using Bayes’ rule as new observations and actions become available, which allows the POMDP to be reformulated as a fully observable MDP, in which each belief state acts as the new state variable. 
Thus, the belief state provides a compact, Markovian representation of the agent’s knowledge about the system, enabling optimal control without the need to store the entire observation-action history. 
The belief state approach to solve \revise{POMDP} is illustrated in Fig.~\ref{fig:methods_compare}a. In model-free methods, the belief is implicitly represented by the hidden state of the RNN, which must be trained jointly with the value and the policy. Alternatively, the belief can be captured explicitly in model-based methods.

Model-based RL methods have been developed to improve sample efficiency and robustness. Here, a model of the environment is either learned or pre-defined. In contrast to model-driven control, this model is used to generate additional samples of the environment for training, such as the Dyna algorithm~\citep{sutton1991DynaIntegratedArchitecture}, 
or to plan over finite horizons, i.e., the agent uses the model to simulate the consequences of candidate action sequences, updating its policy based on the predicted future dynamics~\citep[e.g.,][]{nagabandi2018NeuralNetworkDynamics,NIPS2017_9e82757e,hafner2019DreamControlLearninga}. 
The planning paradigm is closely related to Model Predictive Control (from model-driven control), where the model is used at each time step to optimize a sequence of actions over a finite horizon, with only the first action being implemented before re-planning at the next step. 
Building on this synergy, hybrid approaches have emerged that combine the strengths of model-based and model-free RL. For example,~\citet{schena2024ReinforcementTwinningDigital} proposed an agent that can switch between a model-free RL policy, which learns directly from data, and a model-based policy obtained by adjoint-based optimization, which takes advantage of the predictive model for planning and control. To reduce the computational cost of RL training in high-dimensional environments,~\citet{zeng2022DatadrivenControlSpatiotemporal} develop a reduced model-based strategy in which an autoencoder coupled with a neural ordinary differential equation learns a lower-dimensional representation of the dynamics. To improve data-driven models, physics knowledge has also been incorporated into model-free and model-based RL through symmetry-aware architectures~\citep{zeng2021SymmetryReductionDeep} and physics-informed neural networks~\citep{liu2021PhysicsinformedDynastyleModelbased}.
\\

We propose a framework that transforms the \revise{POMDP} into a MDP with data assimilation, that is, the data-assimilated model-informed RL (DA-MIRL). 
The framework, inspired from observer-based control, is depicted in Fig.~\ref{fig:methods_compare}b, consists of three key components: 
(i) a predictive model of the system's dynamics, 
(ii) sequential data assimilation for real-time state estimation, and 
(iii) reinforcement learning for discovering optimal control policies. 
This approach 
divides the problem of end-to-end training of observer and controller in RL into individual modules. 
%
We combine the ensemble Kalman filter (EnKF) as the state estimator~\citep{evensen2009DataAssimilationEnsemble} with an actor-critic model-free RL method, the Deep Deterministic Policy Gradient (DDPG)\revise{~\citep{lillicrap2016ContinuousControlDeep}}. \revise{Recent studies have explored integrating linear Kalman filters with RL for drone navigation~\citep{marino2024StaticObstaclesIntegrating} and learning a correction for the state estimates of the EnKF in unactuated systems using RL~\citep{hammoud2024DataAssimilationChaotic}. Our framework focuses on spatio-temporally chaotic systems.} The EnKF employs a low-dimensional ensemble of model forecasts to approximate the state covariance, making it scalable to high-dimensional systems such as spatio-temporal chaos.
To model the environment, we consider 
(i) a physics-based low-order model, and (ii) a fully data-driven method---specifically a control-aware echo state network, which is designed here as an efficient architecture that predicts the system's response to control inputs. 
%
The paper is structured as follows. 
In \S~\ref{sec:method}, we define the problem using the notation from RL literature and introduce the building blocks of the framework. In \S~\ref{sec:test_case}, we introduce the Kuramoto-Sivashinsky equation, which is a partial differential equation (PDE) that has spatio-temporal chaos, as the test case for the proposed framework (DA-MIRL). In \S~\ref{sec:results}, we demonstrate the proposed DA-MIRL framework against the model-free RL approach. \S~\ref{sec:conclusion} concludes the paper.

\section{Methodology}\label{sec:method}
We define the time evolution of a controlled system with the discrete-time state space equations
\begin{align}\label{eq:dyn_sys}
\setlength{\arraycolsep}{1.6pt}
\left\{
  \begin{array}{rcl}
  \vect{s}_{k+1} &=& \vect{F}(\vect{s}_k, \vect{a}_k) +\vect{\eta}_k,\\
  \vect{a}_{k} &=& \vect{\pi}(\vect{s}_k) \\
  \end{array}
  \right. \quad \text{for}\;k=0, 1, \dots
\end{align}
where $\vect{s}_k=\vect{s}(t_k) \in \mathcal{S} \subseteq \mathbb{R}^{n_s}$ is the state at time $t_k = k\Delta t$; $\vect{a}_k \in \mathcal{A} \subseteq \mathbb{R}^{n_a}$ is the actuation, i.e., the control input applied to the system at time $t_k$, 
$\vect{\eta}_k\in \mathbb{R}^{n_s}$ encapsulates the aleatoric and epistemic modelling errors; and the initial condition $\vect{s}_0$ is drawn from a normal distribution $\sim \mathcal{N}(\overline{\vect{s}}_0, C_{ss}^0)$. 
The control is applied as a deterministic state-feedback $\vect{\pi}: \mathcal{S} \to \mathcal{A}$ and the dynamics of the system are encapsulated by the operator $\vect{F}: \mathcal{S} \times \mathcal{A} \to \mathcal{S}$, which is Markovian, i.e., 
\begin{equation}\label{eq:Markov_property}
  \vect{s}_{k}\sim\mathcal{P}(\vect{s}_{k} \mid \vect{s}_{k-1}, \vect{s}_{k-2}, \dots, \vect{s}_0, \vect{a}_{k-1}, \vect{a}_{k-2}, \dots, \vect{a}_0) = \mathcal{P}(\vect{s}_{k} \mid \vect{s}_{k-1}, \vect{a}_{k-1})
\end{equation}
where $\mathcal{P}(~\mid~)$ is the transition probability distribution, which in \eqref{eq:Markov_property} describes how likely the system is to transition to a particular state, conditioned on the current state and action. 

In real-world applications, the full system state is often unobservable. Instead, we receive observations from sensors that provide partial and potentially noisy information about the underlying state
\begin{equation}
  \vect{o}_k = \vect{M}(\vect{s}_k) + \vect{\epsilon}_k,
\end{equation}
where $\vect{o} \in \mathcal{O} \subseteq \mathbb{R}^{n_o}$ is the observation vector, $\vect{\epsilon} \in \mathbb{R}^{n_o}$ is the additive measurement noise, and $\vect{M} : \mathcal{S} \to \mathcal{O}$ is the observation operator. We approximate the measurement noise $\vect{\epsilon}$ as Gaussian noise.
Given partial and noisy observations $\vect{o}_k$, our goal is to find the optimal control policy $\vect{\pi}^\star$ that optimizes a certain objective functional, e.g.,~\S~\ref{sec:reward}, constrained by the system's dynamics~\eqref{eq:dyn_sys}. 

\subsection{Reinforcement learning}\label{sec:rl}
First, we define the optimal control problem in an RL setting under the assumption of full state knowledge, i.e., as a MDP. The \revise{MDP} is defined by the state, action, policy and reward spaces, i.e., $(\mathcal{S}, \mathcal{A}, \mathcal{P}, \mathcal{R})$~\citep{sutton2018ReinforcementLearningIntroduction}. 
The optimization objective is defined through the reward function $r(\vect{s}_k, \vect{a}_k): \mathcal{S} \times \mathcal{A} \to \mathcal{R} \subseteq \mathbb{R}$, which assigns a scalar value to each state--action pair, representing the immediate benefit of taking action $\vect{a}$ in state $\vect{s}$ at time $t_k$. 
Hence, given a state $\vect{s}_k$, the agent (i.e., controller) selects an action $\vect{a}_k$ according to a policy $\vect{\pi}$, aiming to maximize the expected return, which is defined as the cumulative reward.
The action-value function~\citep{sutton2018ReinforcementLearningIntroduction}, or $Q$-function, $Q^\pi: \mathcal{S} \times \mathcal{A} \to \mathcal{R}$, evaluates the expected return when taking a specific action $\vect{a}_0$ in state $\vect{s}_0$ at time $t_0$ and subsequently following policy $\vect{\pi}$
\begin{equation}\label{eq:q_function} 
  Q^\pi(\vect{s}_0, \vect{a}_0) = \mathbb{E}\left[{r}(\vect{s}_0,\vect{a}_0) +  \sum_{k=1}^{\infty} \gamma^k r(\vect{s}_k, \vect{\pi}(\vect{s}_k)) \right], \; 
\end{equation}
where $\gamma \in [0, 1]$ is the discount factor that weights the importance of immediate rewards over future ones. 
Due to the Markov property~\eqref{eq:Markov_property}, the optimization problem can be solved recursively leading to the Bellman equation, which is an expectation over the state transition dynamics and policy
\begin{align}\label{eq:q_bellman}
  Q^\pi(\vect{s}_k, \vect{a}_k) &= \mathbb{E} \left[ r(\vect{s}_k, \vect{a}_k) + \gamma Q^\pi\left(\vect{s}_{k+1}, \vect{\pi}(\vect{s}_{k+1})\right) \right]. 
\end{align}
An optimal control policy $\vect{\pi}^\star$ maximizes the action-value function
\begin{equation}\label{eq:action-value}
  Q^\star(\vect{s}, \vect{a}) = \max_{\vect{\pi}}Q^\pi(\vect{s}, \vect{a}), \;\text{for all } \vect{s} \in \mathcal{S} \text{ and } \vect{a} \in \mathcal{A}.
\end{equation}
We employ Deep Deterministic Policy Gradient (DDPG) algorithm~\citep{lillicrap2016ContinuousControlDeep} to solve Eq.~\eqref{eq:action-value} owing to its conceptual and implementation simplicity, as well as to its demonstrated effectiveness in controlling the Kuramoto–Sivashinsky equation~\citep[e.g.,][]{bucci2019ControlChaoticSystems}. 
DDPG is a model-free RL algorithm designed for environments with continuous state and action spaces. Its architecture consists of two neural networks: 
(i) an actor network $\vect{\pi}(\vect{s};\vect{\theta}^\pi)$, i.e., the policy, which maps the state $\vect{s}$ to a continuous deterministic action $\vect{a}$; and 
(ii) a critic network $Q^\pi(\vect{s}, \vect{a};\vect{\theta}^Q)$, which estimates the action-value function associated with the current policy. The actor and critic are parameterized by their respective weights $\vect{\theta}^\pi$ and $\vect{\theta}^Q$.
The actor is trained to select actions that maximize the action-value function~\eqref{eq:action-value}, while the critic is trained to minimize the temporal difference error derived from the Bellman equation~\eqref{eq:q_bellman}. 
Since the underlying system's dynamics are unknown and solving equation~\eqref{eq:q_bellman} becomes intractable in high-dimensional continuous spaces, we approximate the expectation in Eq.~\eqref{eq:q_bellman} via Monte Carlo sampling. 
This sampling is implemented via an experience replay buffer $\mathcal{B}$: 
as the agent interacts with the environment, transition tuples $(\vect{s}_k, \vect{a}_k, r_k, \vect{s}_{k+1})$ are stored in $\mathcal{B}$, and later sampled in batches for training in an off-policy fashion. This approach enhances sample efficiency by allowing multiple updates from past experience. Target networks $\vect{\pi}'$ and $Q^{\pi}{'}$ are updated with an exponential moving average of the learning networks to ensure the stability of the algorithm.
%
The pseudo-code for the network updates are provided in Alg.~\ref{alg:ddpg}.
However, in reality the control problem is a POMDP, because full state observability is often unattainable and only a subset of the full state can be measured---especially for chaotic turbulent fluid systems. With partial observability we account for two additional spaces: the observable $\mathcal{O}$ and likelihood $\mathcal{Z}(\vect{o}_k|\vect{s}_k)$ spaces, where $\mathcal{Z}(\vect{o}_k|\vect{s}_k)$ denotes the probability of observing $\vect{o}_k$ given that the system is in state $\vect{s}_k$~\citep{kaelbling1998PlanningActingPartially}. 
Thus, \revise{POMDPs} are defined in $(\mathcal{S}, \mathcal{A}, \mathcal{P}, \mathcal{R}, \mathcal{O}, \mathcal{Z})$.
%
Although increasing the randomness in the agent via stochastic policies might increase the probability of encountering a good action, conventional RL algorithms to POMDPs yield suboptimal results because the performance of the best stationary stochastic policy can be unboundedly worse than the optimal policy in the underlying MDP~\citep{singh1994LearningStateEstimationPartially}. Another option is keeping a memory of previous actions and observations, which consists of feeding the controller an augmented state or information vector $\vect{I}_k = [\vect{o}_0;\vect{o}_1; \dots \vect{o}_k; \vect{a}_0; \vect{a}_1; \dots \vect{a}_{k-1}]$~\citep{bertsekas2007DynamicProgrammingOptimal}. 
However, when dealing with high-dimensional spatio-temporally chaotic systems, the dimension of $\vect{I}_k$ increases with time, making real-time control computationally intractable. 
A ``sufficient statistic'' encapsulates the necessary information from $\vect{I}_k$, i.e., is Markovian, and has a lower dimension than $\vect{I}_k$ ~\citep{bertsekas2007DynamicProgrammingOptimal}. The belief state~\citep{kaelbling1998PlanningActingPartially}, which expresses the posterior distribution over the state of the system given observations, is a sufficient statistic. For systems with a finite state space, the belief state is a probability vector~\citep{krishnamurthy2016PartiallyObservedMarkov}. For continuous state spaces, the belief becomes a continuous probability density function over the state space. Given a model and observations, the belief, or the state estimate, can be updated sequentially in a Bayesian framework, which we introduce next.

\subsection{State estimation via data assimilation}\label{sec:da}

By assimilating the observations from the environment $\vect{o}$ and the prediction of the full state from the model (the forecast), sequential DA provides an improved prediction on the full state (the analysis) which can then be fed into the actor without the need to modify the RL algorithm. 
%
In a Markovian dynamical systems scenario (c.f., \S~\ref{sec:method}~\eqref{eq:Markov_property}), sequential DA methods leverage Bayesian statistics to provide real-time analysis without full data history. 
The Bayes’ rule defines the analysis (i.e., the posterior of the state estimate) as
\begin{equation}\label{eq:Bayes}
  \underbrace{\mathcal{P}(\vect{s}_k | \vect{o}_k, \matr{F})}_\text{Posterior} \quad\propto \underbrace{\mathcal{P}(\vect{o}_k | \vect{s}_k, \matr{F})}_\text{Observations' likelihood} \, \underbrace{\mathcal{P}(\vect{s}_k, \matr{F})}_\text{Prior}.
\end{equation}
We take the maximum a posteriori (MAP) approach such that the analysis state maximizes the posterior as defined by~\eqref{eq:Bayes}. 
The prior evolves according to the Chapman–Kolmogorov equation, which propagates uncertainty forward in time through the model $\matr{F}$. When the model dynamics are linear and the statistics are Gaussian, the Kalman filter provides an exact solution.
For nonlinear systems, the prior is not analytically tractable, but it can be approximated using ensemble methods based on Monte Carlo sampling. 
Ensemble data assimilation represents the state distribution with a finite set of $m$ samples (the sampling error scales as $\mathcal{O}(m^{-1/2})$). 
The model expectation and uncertainty are approximated by the ensemble mean and covariance, respectively. 
Following~\citep{novoa2024InferringUnknownUnknowns}, we define the augmented state $\vect{\psi} = [\vect{s}; \matr{M}(\vect{s})]$ such that 
\begin{align}
	 \matr{C}_{\psi\psi} = \begin{bmatrix}
				\matr{C}_{ss} & 
        \matr{C}_{s M} \\
				\matr{C}_{M s} & \matr{C}_{MM} \\
			\end{bmatrix}
			&\approx\frac{1}{m-1}\sum^m_{j=1}(\vect{\psi}_j-\bar{\vect{\psi}})\otimes(\vect{\psi}_j-\bar{\vect{\psi}}),
\end{align}
where $\bar{\vect{\psi}}$ is the ensemble mean and $\otimes$ is the dyadic product. 
With this, the ensemble Kalman filter (EnKF) update to the state reads
\begin{align}\label{eq:EnKF}
    \vect{s}_j^\mathrm{a}
  &= 
    \vect{s}_j^\mathrm{f} + 
  \matr{K}{\left[\vect{o}_j - \matr{M}(\vect{s}_j^\mathrm{f})\right]}\\
  \matr{K}&=\matr{C}_{s M}^\mathrm{f}\left(\matr{C}_{oo}+\matr{C}_{MM}^\mathrm{f}\right)^{-1}
\end{align}
where $\matr{K}$ is the Kalman gain, and 
the superscripts `f' and `a' indicate `forecast' and `analysis', respectively. 
To avoid underestimating the uncertainty in the analysis, we perturb the observation $\vect{o}$ such that each ensemble member is updated with a different 
$\vect{o}_j\sim\mathcal{N}(\vect{o}, \matr{C}_{oo})$ , where 
$\matr{C}_{oo}$ is the user-defined observation error covariance~\citep{anderson2001EnsembleAdjustmentKalman}. We define
\revise{\begin{equation}\label{eq:Coo}
  \matr{C}_{oo}(k) = \sigma_o^2\bigg{(\max \big(\big |\matr{M}(\vect{s}(t_k)) \big| \big) \bigg)}^2 \matr{I},
\end{equation}}
where 
$\sigma_o$ is the noise level, 
$\matr{I}$ is the identity matrix, 
the operators \revise{$|\cdot|$} and $\max(\cdot)$ denote element-wise absolute value, and the maximum of a vector, respectively. 
This choice of covariance matrix assumes uncorrelated observations and ensures that the noise is proportional to the observed flow state, which is not statistically stationary when applying control actions to a system before stabilization. 
To further minimize the covariance underestimation, we apply a multiplicative inflation to the ensemble as 
$ 
\vect{s}_j^\mathrm{a} = \overline{\vect{s}}^\mathrm{a} +  \rho_I(\vect{s}_j^\mathrm{a} - \overline{\vect{s}}^\mathrm{a}), 
$
where $\rho_I > 1$ is inflation factor~\citep{evensen2009DataAssimilationEnsemble}. 
A derivation of the Ensemble Kalman filter from a Bayesian perspective is included in App.~\ref{app:EnKF}. 
As illustrated in Fig.~\ref{fig:schematic-DA-RL} the EnKF is an iterative process with two steps: 
\begin{enumerate}
  \item \textit{Forecast step (prior update)}: Forecast the ensemble of states with the model $\matr{F}$ until a time when there are observations $\vect{o}$ available from the environment. This yields the forecast ensemble $\{\vect{s}^\mathrm{f}\}^m_{j=1}$, which is the prior of the states. 
  \item \textit{Analysis step (posterior estimate)}: 
  When observations $\vect{o}$ from the environment are available, we perturb them to create an ensemble $\{\vect{o}\}^m_{j=1}$ and apply the EnKF~\eqref{eq:EnKF}. The EnKF results in the analysis ensemble $\{\vect{s}^\mathrm{a}\}^m_{j=1}$, which is then inflated and used as the new initial condition in~(i).
\end{enumerate}

\subsection{Data-assimilated model-informed reinforcement learning}\label{sec:DA-MIRL}
In model-free RL, the agent acts as a compensator, simultaneously performing both state estimation and control without explicitly modelling the underlying system dynamics. This requires end-to-end training of the observer, which learns a mapping from the observations to a representation of the system's state, and the controller, which learns an optimal policy based on this state. In general, this joint learning problem is highly non-convex and is challenging to optimize. We break this problem down into modules and propose a data-assimilated model-informed reinforcement learning framework, i.e., the DA-MIRL which integrates:
\begin{enumerate}
  \item \textit{Model-free agent}: an off-policy actor-critic RL algorithm for learning the control policy, specifically a DDPG (\S~\ref{sec:rl}). 
  \item \textit{Environment model:} a predictive model of the system's dynamics, i.e., a model of the environment. The model can be either equation-based, or data-driven if the system's equations are unknown or are too expensive for real-time simulation. We discuss two examples of imperfect models in~\S~\ref{sec:imperfect_model}. 
  \item \textit{State estimator}: an ensemble-based data assimilation method for real-time state estimation, specifically the EnKF in this paper. The estimator updates the model predictions using observations from the environment. Then, the full state prediction from the model is fed to the agent, which bypasses the need to modify the implementation of classical RL algorithms for \revise{POMDPs} (\S~\ref{sec:da}). 
\end{enumerate}
Here, we choose the off-the-shelf DDPG and EnKF algorithms to demonstrate the methodology without additional algorithmic complexity. Alternative RL or ensemble DA algorithms (e.g., Soft Actor-Critic or square-root filters) can be implemented for improving the performance whilst keeping the same theoretical background. 

\begin{figure}
  \centering
  \includegraphics[width=.9\linewidth]{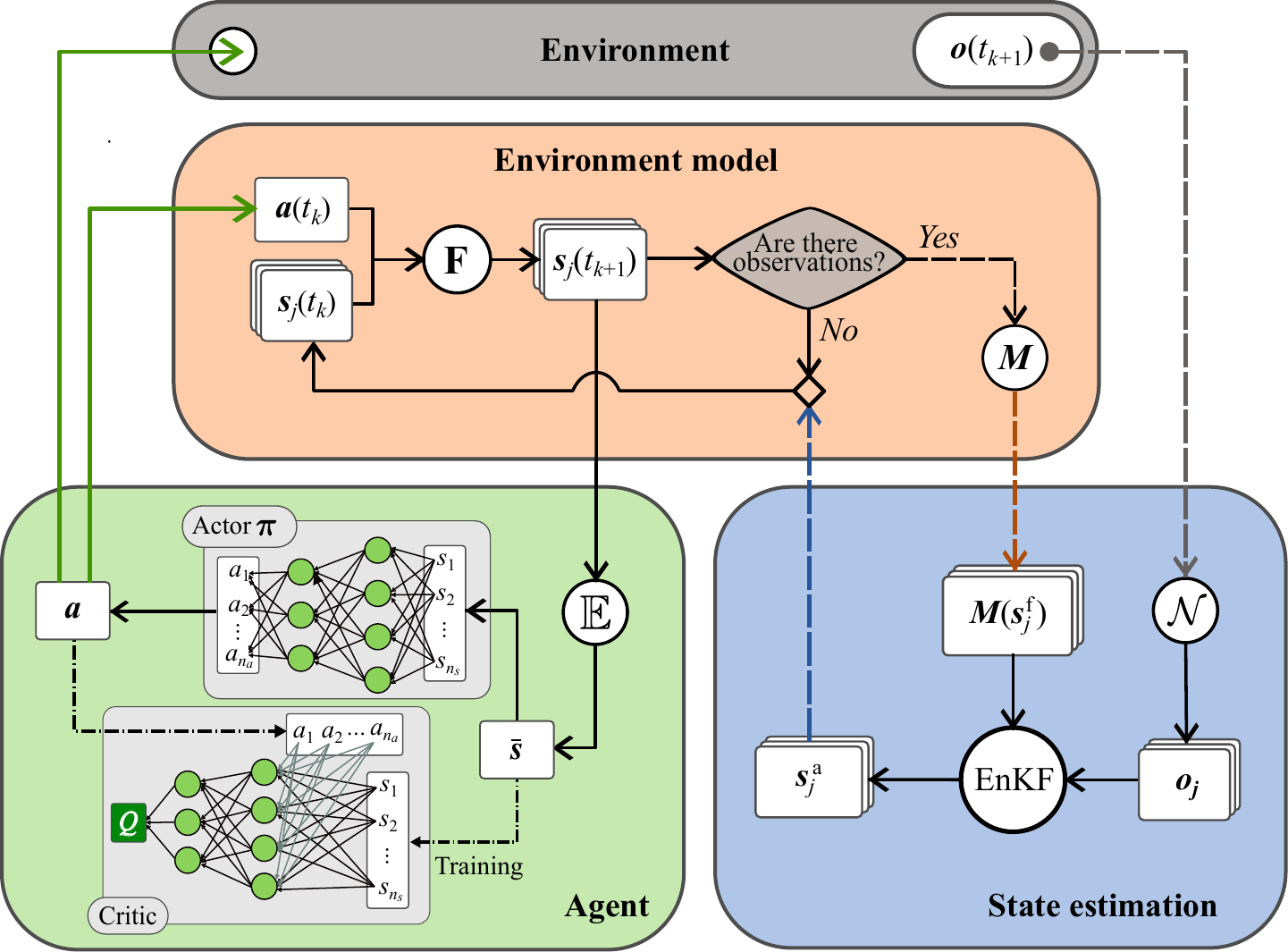}
  \vspace*{-4pt}
  \caption{
  Detailed schematic of the proposed DA-MIRL, which aims to control a partially observed environment (gray). 
  The DA-MIRL integrates three components: model environment (orange), state estimation (blue) and agent (green). 
  The numerical model $\matr{F}$ approximates the environment by forecasting an ensemble of states $\vect{s}_j$ (stacked boxes). 
  If there are no observations, the model runs autonomously, otherwise, the model prediction is updated by the state estimator. 
  The state estimator perturbs the observations $\vect{o}$ and assimilates them with the forecast ensemble via the ensemble Kalman filter (EnKF), which results in the analysis ensemble $\vect{s}^\mathrm{a}_j$. The model $\matr{F}$ is re-initialized with $\vect{s}^\mathrm{a}_j$. 
  The actor-critic agent interacts with both the environment and its model to apply and determine the optimal action at any time $t_k$. 
  The critic, which is active only during training, approximates the $Q$-value function from the state-action pair. 
  The actor (i.e., the policy $\vect{\pi}$) determines the action $\vect{a}$ from the expected value of the full state of the environment. We do not have access to the environment's full state, hence, we feed the expected value given by the model, i.e., the ensemble mean $\bar{\vect{s}}$. 
  }
  \label{fig:schematic-DA-RL}
\end{figure}

The implementation of the DA-MIRL is summarized in Algorithm~\ref{alg:da_mbrl} and illustrated in Fig.~\ref{fig:schematic-DA-RL}. 
The first step is the initialization, both the DDPG (Alg.~\ref{alg:ddpg}) and the ensemble, which is problem-specific (see~\S~\ref{sec:imperfect_model}). 
Then, we propagate the dynamics of the ensemble using the model (forecast step). 
Due to the chaotic effects and modelling uncertainties, the ensemble prediction will de-correlate from the \revise{environment's} true state in time. 
We correct the trajectory of the ensemble every time observations from the environment become available using the EnKF (analysis step) and repeat this loop. After the first analysis step, the state estimate, i.e., the ensemble mean, starts being passed to the agent, which provides the control input. The algorithm (\href{https://github.com/MagriLab/DA-RL}{MagriLab/DA-RL}) is implemented in JAX~\citep{jax2018github}, \revise{leveraging parallelization of the ensemble computations and just-in-time (JIT) compilation for scalability to high-dimensional systems}, and runs on GPU.

\begin{algorithm}
  \caption{DA-MIRL: Data-assimilated model-informed reinforcement learning}
  \label{alg:da_mbrl}
  \begin{algorithmic}[1]
  \State \textbf{Initialize} DDPG, $Q^\pi, \vect{\pi}, Q^{\pi}{'}, \vect{\pi}', \mathcal{B}$ 
  \For{episode = 1 to training episodes}
    \State \textbf{Initialize ensemble}$\{\vect{s}_1, \vect{s}_2, \dots, \vect{s}_m\}$, $\vect{s}_j \sim \mathcal{N}(\bar{\vect{s}}_0, \matr{C}_{ss}^0)$ 
    
    \While {$t_k<$ observation starts}
      \For{each ensemble member $j = 1$ to $m$}
        \State 
        $
        \vect{s}_j \leftarrow F(\vect{s}_j, \vect{0})
        $
      \EndFor
    \EndWhile
    \For {each observations $\vect{o}_k$ in the episode}
      \State \textbf{Perturb} $\vect{o}_{j,k} \sim \mathcal{N}({\vect{o}_k}, \matr{C}_{oo}^k)$
      \For{each ensemble member $j = 1$ to $m$}\Comment {{Analysis step} }
        \State 
        $
        \vect{s}_{j,k}\leftarrow \vect{s}_{j,k} + \matr{K} (\vect{o}_{j,k} - \matr{M}(\vect{s}_{j,k}))
        $~\eqref{eq:EnKF}
        \State $\bar{\vect{s}}_k \leftarrow \frac{1}{m}\sum_j\vect{s}_{j,k}$ 
        \State
        $
        \vect{s}_{j,k}\leftarrow \overline{\vect{s}}_{k} + \rho_I (\vect{s}_{j,k} - \overline{\vect{s}}_{k}) 
        $
      \EndFor
      \While{There are no observations} 
        \State \textbf{Select action} 
        $\vect{a}_k \leftarrow \vect{\pi}(\bar{\vect{s}}_k) + \vect{\epsilon}_{a}$, $ \vect{\epsilon}_{a}\sim\mathcal{N}(\vect{0},\matr{C}_{aa})$
        \For{$j = 1$ to $m$}
        \Comment {Forecast step}
          \State 
          $\vect{s}_{j,{k+1}} \leftarrow \matr{F}(\vect{s}_{j,k}, \vect{a}_k)
          $
        \EndFor
        \State $\bar{\vect{s}}_{k+1} \leftarrow \frac{1}{m}\sum_j\vect{s}_{j,k+1}$
        \State \textbf{Store transition} $(\bar{\vect{s}}_k , \vect{a}_k, r_k, \bar{\vect{s}}_{k+1}) \rightarrow \mathcal{B}$
        \State \textbf{Update networks $Q^\pi, \vect{\pi}, Q^{\pi}{'}, \vect{\pi}'\leftarrow$}Alg.~\ref{alg:ddpg}
        \State 
        $
        k \leftarrow k+1
        $
      \EndWhile
    \EndFor
  \EndFor
  \end{algorithmic}
\vspace*{-4pt} \end{algorithm}

\section{Test case: Kuramoto-Sivashinsky equation}\label{sec:test_case}

The Kuramoto-Sivashinsky (KS) equation is a prototypical model of spatio-temporal chaos, which arises in several physical phenomena such as 
chemical reaction-diffusion systems~\citep{kuramoto1978DiffusionInducedChaosReaction}, 
flame front instabilities~\citep{sivashinsky1980FlamePropagationConditions} and liquid film flows~\citep{sivashinsky1980IrregularWavyFlow}. 
Despite its simple one-dimensional form, the KS partial differential equation (PDE) exhibits rich nonlinear dynamics and spatio-temporal chaos, making it a benchmark system for studying control of chaotic flows~\citep[e.g.,][]{gomes2017StabilizingNontrivialSolutions,shawki2020FeedbackControlChaotic,bucci2019ControlChaoticSystems}. Including a forcing term to implement controls, the equation is described by 
\begin{equation}\label{eq:pde_ks}
  \frac{\partial u}{\partial t} + \frac{\partial^2 u}{\partial x^2} + \frac{\partial^4 u}{\partial x^4} + u\frac{\partial u}{\partial x} = f(x,t),
\end{equation}
where $u(x, t): [0, L) \times [0, \infty) \rightarrow \mathbb{R}$ denotes the velocity on a periodic domain of length $L$, i.e., $u(x, t) = u(x+L, t)$, and $f$ is the forcing term (here, the control input). Equivalently, Eq.~\eqref{eq:pde_ks} can be scaled to $2\pi$-periodic domains, which introduces the viscosity parameter $\nu = (2\pi/L)^2$ multiplying the fourth order term~\citep{gomes2017StabilizingNontrivialSolutions}. 
We numerically solve the PDE~\eqref{eq:pde_ks} as follows. 
First, we discretize $u(x,t)$ using the Fourier expansion for even $n_f$ modes
\begin{equation}\label{eq:inverse_fourier}
  u(x,t) \approx \frac{1}{n_f}\left(c_0(t) + c_{n_f/2}(t)\exp\left(i \frac{\pi n_f}{L} x\right) + \sum_{l=1}^{n_f/2-1}c_{l}(t)\exp\left(i \frac{2\pi l}{L} x\right) + c.c. \right),
\end{equation}
where we retain the Fourier coefficients corresponding to the non-redundant positive wavenumbers $\vect{c}^\top=(c_0, \dots, c_{n_f/2})$ and ensure real-valuedness via the complex conjugate. Second, we propagate the dynamics of the Fourier coefficients,
\begin{equation}\label{eq:spectral_ks}
  \frac{dc_l}{dt} = (\kappa_l^2 - \kappa_l^4)c_l(t) - \frac{1}{2}i\kappa_l\sum_{\substack{p+q=l}}c_p(t)c_q(t) + f_l(t),
\end{equation}
where $\kappa_l = \frac{2\pi l}{L}$ is the wavenumber
and $f_l(t)$ is the Fourier coefficient of the forcing term at the wavenumber $\kappa_l$ (see~\S~\ref{sec:control_strategy}). 
Third, the set of equations~\eqref{eq:spectral_ks} is integrated using a semi-implicit third-order Runge-Kutta scheme~\citep{kar2006SemiImplicitRungeKutta}.

%
The nonlinear advection term $u \partial u/\partial x$ is responsible for the transfer of energy between different scales and, 
as the viscosity parameter decreases $\nu<1$ (or equivalently $L>2\pi$), the stable solution $\vect{u}=\vect{0}$ becomes unstable and 
bifurcates, giving rise to travelling wave solutions. Further decrease in $\nu$ leads to a series of period-doubling bifurcations leading to chaotic dynamics. This chaotic regime is characterized by complex, aperiodic dynamics~\citep{papageorgiou1991RouteChaosKuramotoSivashinsky}.

We test the DA-MIRL framework using a twin experiment approach. 
Thus, we define a numerically simulated \textit{true} environment of the KS equation (\S~\ref{sec:environment}) which is assumed unknown, and we approximate the truth with approximate models of the environment (\S~\ref{sec:imperfect_model}). 
The same control input is applied to both the true and model environments (\S~\ref{sec:agent}), and the performance of the DA-MIRL  is evaluated by comparing model predictions with the true solution.

\subsection{The environment}\label{sec:environment}

The state of the environment (the truth) is the solution of the KS in the physical domain, resolved using $n_f^{true} = 64$ Fourier modes, i.e., 
\begin{equation}\label{eq:true_state}
   \vect{u}_k^{true}(\vect{x}_s)= \mathcal{F}^{-1}(\vect{c}^{true}_k) 
\end{equation} 
where 
$\vect{x}_s^\top = \left(0, 1, \dots, 63\right)L/64$ and 
$\mathcal{F}$ denotes the Fourier transform. 
This true state is assumed unknown. Instead, the sensors provide observations $\vect{o}_k$, which are noisy discrete samples of~\eqref{eq:true_state} 
\begin{equation}
  \vect{o}_k \coloneqq \vect{u}_k^{true}(\vect{x}_o) + \vect{\epsilon},
\end{equation}
where $\vect{x}_o^\top = \left(0, 1, \dots, (n_o-1)\right)L/n_o$ 
are the sensors' locations, $\vect{\epsilon}$ is the observation noise (which we assume Gaussian in~\S~\ref{sec:da}), and the state is fully observed with $n_o=64$. 
%
\subsection{The environment model}\label{sec:imperfect_model}
Since true state of the system is assumed unknown, we use surrogate models to approximate the underlying dynamics of the environment. We test the DA-MIRL framework on a physics-based model (\S~\ref{sec:model_truncated}) and a data-driven (\S~\ref{sec:model_esn}) model. \revise{The framework is applicable across the spectrum of modelling choices, from purely physics-based to purely data-driven, where hybrid modelling approaches can also be explored.}
The model prediction of the true state at time $t_k$, denoted by $\hat{\vect{u}}_k$, is corrected with observations $\vect{o}_k$, as detailed in \S~\ref{sec:da}. 

\subsubsection{Physics-based model: truncated Fourier model}
\label{sec:model_truncated}
We first consider a 
physics-based low-order model of the KS equation by truncating the Fourier series to a lower number of modes than used to create the environment (\S~\ref{sec:environment}). This is, $n_f<n_f^{true}=64$. 
This scenario is encountered in reality when we have an accurate model, but because of computational constraints, we cannot fully resolve it and as a result we have a coarse-grained model.
The truncation discards the effects of higher wavenumber modes, which are nonlinearly coupled with the smaller wavenumber \revise{modes}. 
The truncation leads to aliasing when projecting the forcing $f(x,t)$, which degrades further the model performance. 
To circumvent this issue, we first resolve the forcing on a finer spatial grid and then truncate it to determine the forcing $f_l(t)$ in Eq.~\eqref{eq:spectral_ks} (c.f. \eqref{eq:gaussian_ks}). 

The augmented state vector of the truncated Fourier model, as defined \S~\ref{sec:da} is
\begin{equation}\label{eq:psi_fourier}
  \vect{\psi}_j =
  \begin{bmatrix}
    \vect{s}_j\\
    \matr{M}(\vect{s}_j)
  \end{bmatrix}
  \coloneqq
  \begin{bmatrix}
    \vect{c}_j\\
    \mathcal{F}^{-1}(\vect{c}_j(t_k))(\vect{x}_o)
  \end{bmatrix}.
\end{equation} 
for each ensemble member $j=1, \dots, m$, 
and the expected reconstructed state is given by 
\begin{equation}
  \bar{\vect{s}}_{RL}=\frac{1}{m}\sum \hat{{\vect{u}}}_j \coloneqq \frac{1}{m}\sum\mathcal{F}^{-1}(\vect{c}_j), 
\end{equation}
which is the input to the actor-critic networks. This mapping to the real-domain defined by the inverse Fourier transform~\eqref{eq:inverse_fourier} avoids complex-valued inputs to the actor-critic.
Each ensemble member is initialized by sampling from a complex Gaussian distribution centered at $\vect{c}(t_0)=\vect{c}_0$. 
The real and imaginary components are treated independently, i.e., 
$\mathrm{Re}(\vect{c}_j(t_0)) \sim \mathcal{N}\left(\mathrm{Re}(\vect{c}_0), \sigma_{0}^2 \mathrm{diag}(|\mathrm{Re}(\vect{c}_0)|)\right)$ 
and 
$\mathrm{Im}(\vect{c}_j(t_0)) \sim \mathcal{N}\left(\mathrm{Im}(\vect{c}_0), \sigma_{0}^2 \mathrm{diag}(|\mathrm{Im}(\vect{c}_0)|)\right)$, 
where $\sigma_{0}$ is the initial uncertainty scaling.
This complex Gaussian structure respects Fourier transform symmetries, assumes independence between Fourier modes and preserves the spectral energy distribution. 




\subsubsection{Data-driven model: Control-aware Echo State Network} 
\label{sec:model_esn}

\begin{figure}
  \centering
  \includegraphics[width=\linewidth]{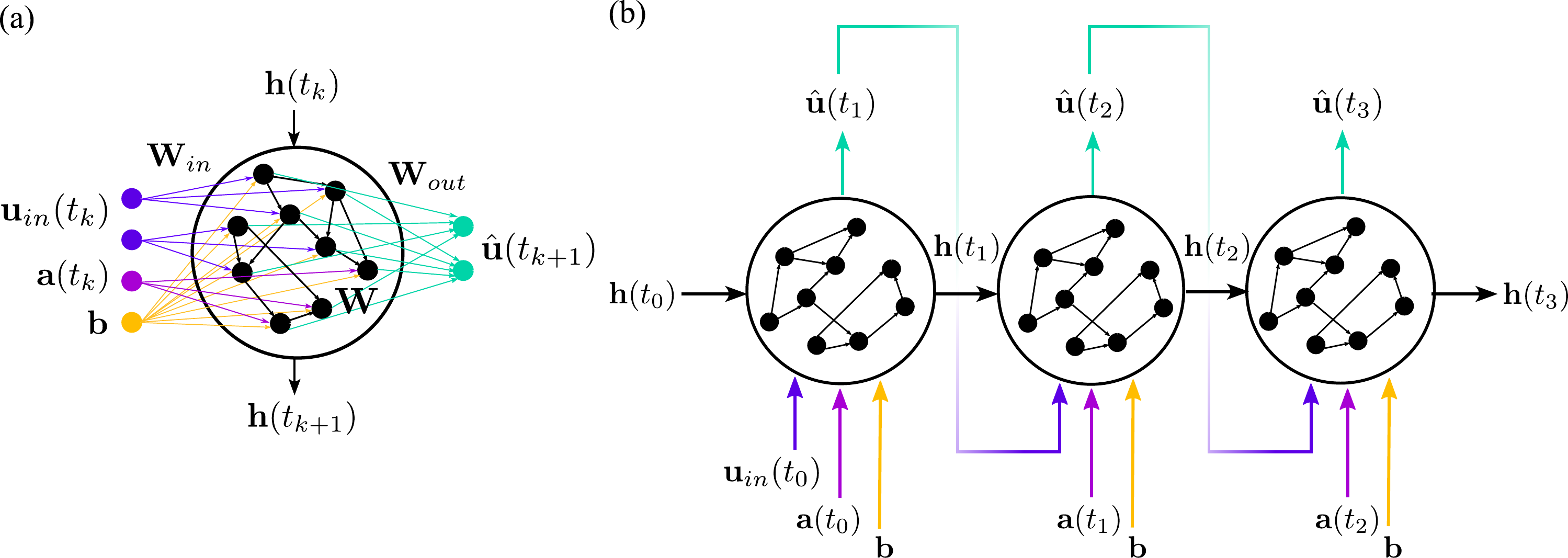}
  \vspace*{-4pt}
  \caption{Control-aware echo state network. (a) Compact schematic of the architecture, and (b) pictorial forecast of the network with one open-loop followed by two closed-loop steps.}
  \label{fig:esn_schematic}
\end{figure}

Second, we adopt a data-driven approach to approximate the dynamics of the true state. 
We seek a model that 
(i) captures temporal dependencies in the data, 
(ii) generalizes across different dynamical regimes, and 
(iii) has a state-space representation suitable for state-feedback control, i.e., is control-aware.
Echo State Networks (ESNs) are a class of recurrent neural networks which satisfy (i)--(ii) by encoding time-series data into a high-dimensional hidden state (the reservoir), which acts as a numerical memory and, for unactuated systems, ESNs are universal approximators~\citep{grigoryeva2018EchoStateNetworks}. To address (iii), we propose a control-aware ESN~\citep{ozan2025DataAssimilatedModelBasedReinforcement,racca2023ControlawareEchoState}, which extends the standard ESN architecture by augmenting the input layer with action signals. The architecture of the control-aware ESN is illustrated in Fig.~\ref{fig:esn_schematic}. 
Further, ESNs are attractive for real-time applications as their training consists of solving a linear system, which is computationally cheaper than backpropagation-based methods and has one global minimum. 
The state-space formulation of the control-aware ESN is 
\begin{equation}\label{eq:control_esn_step}
  \vect{h}(t_{k+1}) = (1-\alpha)\vect{h}(t_k) + \alpha\tanh(\matr{W}_{in}\yintilde(t_k)+\matr{W}\vect{h}(t_k)+\vect{b}),
\end{equation}
where 
$\vect{h} \in \mathbb{R}^{n_h}$ is the reservoir state, 
$\vect{b}\in \mathbb{R}^{n_h}$ is the input bias vector, whose entries are sampled from $\mathcal{U}(-1,1)$, is introduced to break the anti-symmetry of the $\tanh(\cdot)$ element-wise operation, 
and 
$\yintilde\in \mathbb{R}^{(n_u+n_a)}$ is the augmented input, defined as 
\begin{equation}
  \yintilde(t_k) = [(\yin(t_k)- \overline{\vect{U}})\oslash \vect{\sigma}_{U}; \vect{a}(t_k)], 
\end{equation}
i.e., the concatenation of the raw input state $\yin$ 
(standardized component-wise by the 
temporal mean $\overline{\vect{U}}$ 
and standard deviation $\vect{\sigma}_{U}$ 
of the training data $\matr{U} \in \mathbb{R}^{n_u \times N_{t}}$, where $\oslash$ indicates element-wise division) 
and the control actions $\vect{a}$ (not normalized since $a_i\in[-1,1]$). 
The reservoir dynamics~\eqref{eq:control_esn_step} are defined by the input matrix $\matr{W}_{in}\in \mathbb{R}^{n_h \times (n_u+n_a)}$, which projects the augmented input $\yintilde$ into the reservoir state space, and the state matrix $\matr{W}\in \mathbb{R}^{n_h \times n_h}$, which governs the internal connections between the reservoir states. 
Each row of $\matr{W}_{in}$ contains a single randomly placed non-zero element sampled from $\sim \mathcal{U}(-1,1)$ and scaled by the factor $\xi_{u}$ or $\xi_{a}$, depending whether the element acts on an input state or action input. This yields the structure 
$\matr{W}_{in} =[\xi_{u}\tilde{\matr{W}}_{in}^{u}~~\xi_{a}\tilde{\matr{W}}_{in}^{a}]$, where 
$\tilde{\matr{W}}_{in}^{u}$ and $\tilde{\matr{W}}_{in}^{a}$ are the sub-matrices associated to the state and action components of $\yintilde$. 
\revise{The state matrix is defined as $\matr{W} = \rho \tilde{\matr{W}}$, where $\tilde{\matr{W}}$ is an Erd\H{o}s-Renyi matrix with connectivity $n_{conn}$, i.e., each row has an average of $n_{conn}$ non-zero elements drawn from a uniform distribution $\sim \mathcal{U}(-1,1)$ and scaled such that $\tilde{\matr{W}}$ has unity spectral radius and the hyperparameter $\rho$ defines the spectral radius of $\matr{W}$.}
The matrices $\matr{W}$ and $\matr{W}_{in}$ are randomly generated and fixed at initialization, and the readout layer of the ESN is defined via training. The readout has the linear form
\begin{equation}\label{eq:control_esn_readout}
  \hat{\vect{u}}(t_{k+1}) = \matr{W}_{out}[\vect{h}(t_{k+1}); 1], 
\end{equation}
where $\hat{\vect{u}} \in \mathbb{R}^{n_u}$ is the ESN prediction, 
the constant $1$ incorporates an output bias, 
and $\matr{W}_{out} \in \mathbb{R}^{n_u \times (n_h+1)}$ is the output matrix, whose weights are determined through training. 
We train the control-aware ESN offline on the full state data from the system generated by random actuations. 
The training and validation procedures consisting of solving a Ridge regression problem~\citep{jaeger2004HarnessingNonlinearityPredicting} and validation with Bayesian optimization of the ESN hyperparameters are detailed in App.~\ref{app:training}. 
As illustrated in Fig.~\ref{fig:esn_schematic}(b), the ESN can operate in two configurations: open and closed loop. The two configurations differ only in $\yin$: if there is data available, the network is forecast in open loop (i.e., during training and initialization), otherwise, the prediction is used as input, i.e., $\yin(t_k) = \hat{\vect{u}}(t_k)$, this is, the network propagates autonomously in closed loop. 

The control-aware ESN is used as the surrogate model of the environment. During the DA-MIRL deployment, the ESN runs the ensemble in closed loop between observations, and we perform a single open-loop step at with the analysis ensemble as input. 
Following~\citep{novoa2025OnlineModelLearning}, we update the ESN's reservoir state via the EnKF, i.e., we define 
\begin{equation}\label{eq:psi_esn}
  \vect{\psi}_j =
  \begin{bmatrix}
    \vect{s}_j\\
    \matr{M}(\vect{s}_j)
  \end{bmatrix}
  \coloneqq
  \begin{bmatrix}
    \vect{h}_j\\
    \matr{W}_{out}[\vect{h}_j; 1](\vect{x}_o)
  \end{bmatrix}.
\end{equation} 
for each ensemble member $j=1, \dots, m$, 
and the expected reconstructed state is 
\begin{equation}
  \bar{\vect{s}}_{RL}=\frac{1}{m}\sum \hat{{\vect{u}}}_j \coloneqq \frac{1}{m}\sum \matr{W}_{out}[\vect{h}_j; 1], 
\end{equation}
which is the input to the actor-critic networks. 
To initialize the ensemble of reservoir states, we first initialize an ensemble of $\vect{u}_j(t_0)$ using the same sampling strategy as the Fourier model in \S~\ref{sec:model_truncated} for comparability of the results. 
To initialize $\vect{h}_j$, we input $\vect{u}_j(t_0)$ to~\eqref{eq:control_esn_step} in open loop repeatedly for the length of the reservoir initialization, i.e., $n_{wash}=100$ time steps. 
Using a single instance of the input state for the washout instead of a sequence may introduce additional error, however, this error is mitigated by the data assimilation process.
%
During the offline training of the ESN, we assumed that the full-state information is available. This choice is similar to the concept of privileged information methods in RL~\citep[e.g.,][]{baisero2022AsymmetricDQNPartially}, which assume access to full state information provided by simulators during offline training and only partial observations during online deployment. 
The training of the ESN (or RNNs in general) on partial observations is a considerably more difficult task than training on the full state~\citep{gupta2024PredictabilityWeaklyTurbulent}, which can be addressed for example by physics-informed approaches~\citep[e.g.,][]{doan2020LearningHiddenStatesa,ozalp2023ReconstructionForecastingStability}. If such methods are successfully applied, and the ESN output remains a partial observation of the underlying system, the controller can instead operate on the reservoir state. \revise{To tackle high-dimensional problems, strategies such as employing parallel ESNs for different parts of the state~\citep{pathak2018ModelFreePredictionLarge} or coupling the ESN with an autoencoder for order reduction~\citep{racca2023PredictingTurbulentDynamics} can also be integrated in the DA-MIRL framework.}
\subsection{The agent}\label{sec:agent}
Our objective is to solve the optimal control problem to find the control actions that stabilize the KS system starting from a solution in the chaotic attractor. 
To this aim, we implement a DDPG actor-critic agent (\S~\ref{sec:rl}). 
The agent interacts with both the environment (\S~\ref{sec:environment}) and its model (\S~\ref{sec:imperfect_model}), either the truncated Fourier model or the control-aware ESN, enabling simultaneous policy improvement and model adaptation. 
\subsubsection{Actuation strategy}\label{sec:control_strategy}
We define the control forcing to the KS equation~\eqref{eq:pde_ks} using $n_a$ 
 spatially distributed actuators, which are modelled as Gaussian basis functions
\begin{equation}\label{eq:gaussian_ks}
  f(x, t) = \sum_{i = 1}^{n_a}a_i(t) \exp\left(-\frac{(x-x_{a}^i)^2}{2\sigma_a^2}\right),
\end{equation}
$\vect{a}^\top = \left(a_1, a_2, \dots, a_{n_a}\right)$ are the control actions, with $a_i \in [-1,1]$; 
$\vect{x}_a^\top = \left(x^1_{a}, x^2_{a}, \dots, x^{n_a}_{a}\right)$ denote actuator locations
and $\sigma_a$ is the width of the actuator, i.e., the spatial influence of the action. 
%
The agent interacts with the environment (and its models) in a zero-order hold (ZOH) approach, i.e., after each control action, the most recent value is kept unchanged until a new action becomes available. The ZOH approach results in a step-like signal between updates. 

\subsubsection{Reward function}\label{sec:reward}

To drive the system to the trivial, i.e., zero, solution whilst spending minimal actuation power, we define the reward as a combination of the $\ell_2$-norms of the state and the action
\begin{equation}\label{eq:reward_ks}
  r(\vect{u}_k,\vect{a}_k) = -\left(\frac{1}{\sqrt{n_f}}||\vect{u}_{k+1}||_2 + \lambda_a||\vect{a}_k||_2\right),
\end{equation}
where $\lambda_a$ penalizes excessive actuation and 
the state $\ell_2$-norm is scaled by $\sqrt{n_f}$ such that the reward is comparable for models with different number of Fourier modes. The reward function in this setup~\eqref{eq:reward_ks} depends on the full state $\vect{u}$. However, in practical applications, only partial measurements of $\vect{u}$ is available. 
In POMDP settings, it is common to assume that the reward is observable, even when the full state is not. In the DA-MIRL approach, we find that providing the critic with a reward computed from the true (but unobserved) state, whilst relying on the estimated state for value prediction, destabilizes its training if a discrepancy between the true state and its estimate exists. Therefore, we also estimate the reward using the model, i.e., we compute $r(\hat{\vect{u}}_k,\vect{a}_k)$, for both feasibility and consistency. 
%

\subsubsection{Performance metrics}\label{sec:metrics}
\paragraph{Episode return.}
We define the episode return as the sum of non-discounted rewards in an episode after the observation starts
\begin{equation}
 R = \sum_{k = k_{start}}^{k_{start}+k_{ep}}r(\vect{u}_k,\vect{a}_k),
\end{equation}
which is computed using the full state vector $\vect{u}_k$ from the true system, i.e., the environment, The index $k_{ep}$ denotes the episode length and the index $k_{start}$ denotes when the observations start being assimilated after the ensemble is initialized.
\paragraph{Actor and critic losses.}
The actor and critic networks are trained simultaneously using randomly sampled batches of size $n_b$ from the experience replay buffer at each time step. 
The critic network is trained 
by minimizing the loss function~\citep{lillicrap2016ContinuousControlDeep,sutton2018ReinforcementLearningIntroduction}
\begin{equation}\label{eq:L_Q}
L_Q = \frac{1}{n_b} \sum_{i = 1}^{n_b} \left(\delta^{(i)}\right)^2, 
\end{equation}
where $\delta^{(i)}$ is the temporal difference error for each batch $i = 1,2, \dots, n_b$, which is defined as
\begin{equation}
 \delta^{(i)} = r_k^{(i)} + \gamma Q^{\pi}{'}\left(\vect{s}_{k+1}^{(i)}, \vect{\pi}'\left(\vect{s}_{k+1}^{(i)}\right)\right) - Q^\pi\left(\vect{s}_k^{(i)}, \vect{a}_k^{(i)}\right), 
\end{equation}
Although the critic is trained to enforce the Bellman optimality condition~\eqref{eq:q_bellman}, the actor is trained to maximize the return from the learned $Q$-function 
\begin{equation}\label{eq:L_pi}
L_\pi = -\frac{1}{n_b} \sum_{i = 1}^{n_b} Q^\pi\left(\vect{s}_k^{(i)}, \vect{\pi}\left(\vect{s}_k^{(i)}\right)\right).
\end{equation}
The losses~\eqref{eq:L_Q} and~\eqref{eq:L_pi} are differentiated with respect to the critic and actor weights (see \revise{Alg.~\ref{alg:ddpg}}). The policy loss is minimized via gradient descent using the policy gradient. 

\section{Results}\label{sec:results}
We demonstrate the DA-MIRL framework on the KS equation. The objective is threefold: 
(i) to construct a predictive model that accurately forecasts the dynamics, 
(ii) to perform real-time state estimation from partial and noisy observations via data assimilation, and 
(iii) to learn an optimal control policy based on the state estimates. 
The training is composed of three stages of episodes: 
(1) randomly actuated, where actions are randomly sampled to create experiences for the replay buffer, 
(2) exploration-learning, where the RL agent explores the solution space and the weights of the actor-critic are updated, and 
(3) evaluation episodes, where the performance of the RL agent is evaluated to determine the best actor-critic weights. 
We apply model-free RL to the KS equation in \S~\ref{sec:results_model_free} and then compare these results with the proposed DA-MIRL framework in \S~\ref{sec:results_DAMBRL}. 
The test cases in \S~\ref{sec:results_model_free}-\S~\ref{sec:results_DAMBRL} are performed on the KS equation with $\nu=0.08$ ($L\approx22$). In the unactuated case, this is a chaotic regime, where the characteristic timescale of the divergence of the nearby trajectories, known as the Lyapunov time (LT), is approximately equal to 25 time units in Eq.~\eqref{eq:pde_ks} or 500 time steps.
To assess robustness, additional results in other chaotic regimes are shown in \S~\ref{sec:results_esn}. 
The training details are included in App.~\ref{app:training}.

\subsection{Model-free reinforcement learning }\label{sec:results_model_free}

\begin{figure}
  \centering
  \includegraphics[width=\textwidth]{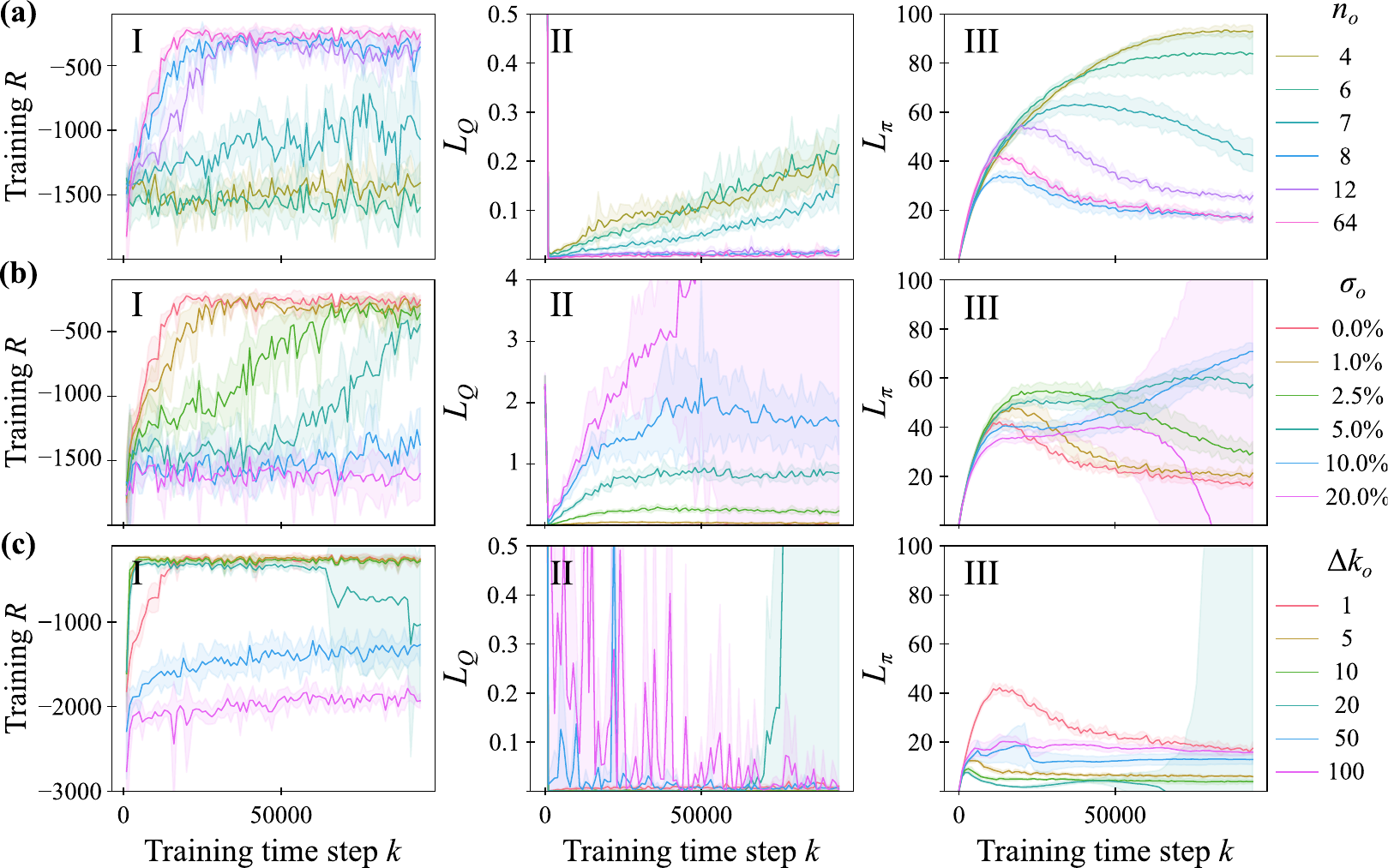}
  \vspace*{-4pt}
  \caption{
  Model-free RL: training performance for
  varying 
  (a) number of sensors; 
  (b) levels of observation error level, $\vect{\sigma}_{o}$; and 
  (c) frequency of observations ($\Delta k_o$).
  Mean and standard deviation of the performance metrics across 5 different runs: 
  (I) return $R$, (II) critic loss $L_Q$, (III) policy loss $L_\pi$. 
  Default parameters: 
  $n_a=8$, $n_o=64$, $\sigma_o=0.0\%$, $\Delta k_o=1$.
  }\label{fig:mf_all}
\end{figure}

We analyze the effectiveness of the DDPG algorithm (without state estimation) for stabilizing the KS system. We examine the agent's performance metrics (\S~\ref{sec:metrics}) for varying 
(i) the number of sensors, 
(ii) the location of sensors and actuators, 
(iii) the level of observation noise, and 
(iv) the frequency of observation and actuation (Fig.~\ref{fig:mf_all}). 
Figure~\ref{fig:mf_all}a shows that the model-free agent fails to stabilize the system with $n_o<8$ sensors, which is consistent with previous works~\citep{bucci2019ControlChaoticSystems, weissenbacher2025ReinforcementLearningChaotic}. 
We find that the observability is further affected by the position of actuators and sensors---staggered actuators/sensors (i.e., $\vect{x}_{a} \neq \vect{x}_{o}$) yield slower convergence 
whereas optimal performance is achieved when $\vect{x}_{a} = \vect{x}_{o}$. 
Although $n_o = 8$ fulfills the Nyquist-Shannon sampling criterion for the first four dominant Fourier modes, this is not a sufficient condition to determine the observability threshold. \revise{Empirical studies~\citep{gupta2024PredictabilityWeaklyTurbulent,weissenbacher2025ReinforcementLearningChaotic} suggest that, in chaotic systems, this threshold is related to the dimension of the inertial manifold, which is coincidentally equal to 8 for the $\nu = 0.08$ regime~\citep{cvitanovic2010StateSpaceGeometry}, or the number of sensors needed for chaos synchronization. But, formalizing an observability condition remains an open question, as the observability may differ for the actuated system and depend on the sensor positions.}
Next, Fig.~\ref{fig:mf_all}b quantifies the impact of observation noise on policy learning in the fully-observable case, i.e., $n_o = 64$. Although the agent stabilizes for noise levels for $\sigma_o \leq 5.0\%$, \revise{higher variances} ($\sigma_o \geq10\%$) critically degrade the policy learning and destabilizes the approximation of the value function, as evidenced by $L_Q$ divergence in Fig.~\ref{fig:mf_all}c.II. Noise corrupts the state in the Bellman updates and introduces bias in $Q$-value function. Thus, the agent must sample more trajectories to average out these biases. Further, the noise inflates the variance of rewards, which may cause over-exploration due to environmental stochasticity or under-exploitation, as optimal policies can be masked by noise. 
Lastly, 
we investigate the effect of the observation frequency $\Delta k_o^{-1}$ in the fully-observable noise-free case in Fig.~\ref{fig:mf_all}(c). 
We vary the observation interval between $\Delta k_o \in [1, 100]$, i.e., under the Lyapunov time for the system. 
With intervals of 5–10 steps, we obtain similar performance as interacting with the environment every time step, and faster convergence. The decrease in sample complexity can be explained by the exploration of more of the phase space with \revise{fewer} samples, or that the ZOH strategy simplifies the training. 
Although the performance drops as $\Delta k_o>20$, the actor-critic losses decrease over time, in contrast to the sparse or noisy measurement cases.

To summarize, the performance of the model-free DDPG algorithm depends on the number and placement of sensors, the noise level, and the frequency of observation/actuation. These factors impact the stability RL agent training, which is evidenced by the divergence of the actor-critic losses. In the next section, we apply the DA-MIRL framework to overcome these limitations, which arise in real-life applications.

\subsection{Data-assimilated model-informed reinforcement learning }\label{sec:results_DAMBRL}

We evaluate the DA-MIRL framework using the two models described in \S~\ref{sec:imperfect_model} to approximate the higher-fidelity environment. The results for the truncated Fourier model are presented in \S~\ref{sec:results_fourier}, and those for the control-aware ESN are shown in \S~\ref{sec:results_esn}. 
Tab.~\ref{tab:notation} summarizes the notation of the model and the reference environment quantities.
As outlined in \S~\ref{sec:DA-MIRL}, we employ an ensemble to infer the first two statistical moments. We find that an ensemble size of $m = 50$ offers a good trade-off between computational efficiency and performance---increasing the ensemble size from $m = 10$ to $m = 50$ yields an approximate $50\%$ improvement in the mean episode return after convergence, whereas increasing from $m = 50$ to $m = 100$ gives only a marginal improvement of $\sim5\%$. 
We update the ensemble prediction by assimilating sparse and noisy observations from the environment. The observations are unbiased samples from the environment with a noise level $10\%$, i.e., 
$\sigma_o = 0.1$ in \eqref{eq:Coo}. \revise{The EnKF assumes that the environment’s model and the observation noise are unbiased. To overcome biased state estimates that can occur due to modelling errors or sensor drifts, a bias-aware data assimilation method~\citep{novoa2024InferringUnknownUnknowns} can be employed.} Observations are assimilated at intervals of $\Delta k_o = 10$, i.e., the observation frequency is $\Delta k_o^{-1} = 1/10$. 
The parameters are listed in Tab.~\ref{tab:default}. 
\begin{table}[]
\caption{Summary of the relation between the environment quantities and their model approximations. }
\label{tab:notation}
\begin{tabular}{l|c|c|c|c|}
\cline{2-5}
\multicolumn{1}{c|}{} & \multicolumn{1}{c|}{\multirow{3}{*}{Environment}} & \multicolumn{3}{c|}{Environment model $\vect{F}$} \\ \cline{3-5} 
& & 
General form
 & 
 Truncated Fourier
 &
 Control-aware ESN
\\ \hline
\multicolumn{1}{|l|}{Full state} & $\vect{u}^{true} $ (unknown) & 
$\hat{\vect{u}}$ & $\mathcal{F}^{-1}(\overline{\vect{c}})$ & $\matr{W}_{out}\left[\overline{\vect{h}};1\right]$ \\ \hline
\multicolumn{1}{|l|}{Ensemble} & $-$ & $\vect{s}_j$& $\vect{c}_j$ & $\vect{h}_j$ \\ \hline
\multicolumn{1}{|l|}{Observable} & 
$\vect{o} = \vect{u}^{true}(\vect{x}_o) + \vect{\epsilon} $ 
& $\matr{M}(\vect{s}_j)$ & $\mathcal{F}^{-1}(\vect{c}_j)(\vect{x}_o)$ & $\matr{W}_{out}[\vect{h}_j;1](\vect{x}_o)$\\ \hline
\end{tabular}
\vspace*{-4pt} 
\end{table}

\begin{table}[]
\caption{Configuration for Figs.~\ref{fig:mb_fo_1}-\ref{fig:mb_esn_nu}
unless stated otherwise.
}\label{tab:default}
\centering
\begin{tabular}{lll}
Parameter & Symbol & Value \\ \hline
Number of actuators & $n_a$ & 8 \\
Actuators' locations & $\vect{x}_a$ & $(0,L/8,\dots,7L/8)^\top$ \\
Number of sensors & $n_o$ & 4 \\
Sensors' locations & $\vect{x}_o$ & $(0,L/4,L/2,3L/4)^\top$ \\
Observation interval & $\Delta k_o$ & 10 \\
Control interval & $\Delta k_a$ & 1 \\
Environment Fourier modes & $n_f^{true}$ & 64 \\
Truncated Fourier modes & $n_f$ & 16 \\
Reservoir size & $n_h$ & 1000 \\
Ensemble size & $m$ & 50 \\
Ensemble inflation & $\rho_I$ & 1.02 \\
KS viscosity & $\nu$ & 0.08 \\
\end{tabular}
\vspace*{-4pt} 
\end{table}

\subsubsection{Truncated Fourier model}\label{sec:results_fourier}

\begin{figure}
  \centering
  \includegraphics[width=\textwidth]{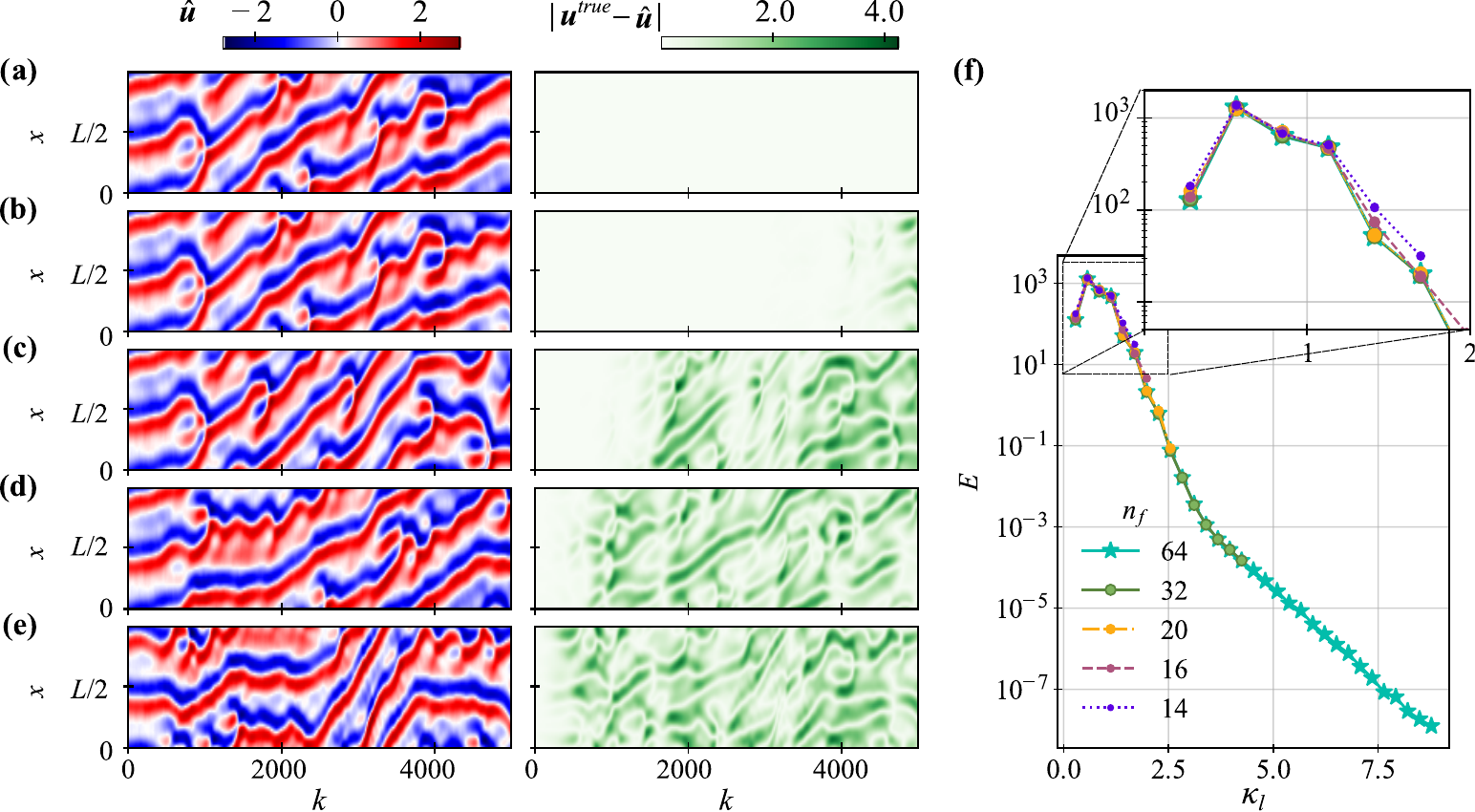}
  \vspace*{-4pt}
  \caption{Numerical simulations of the KS equation with varying number of Fourier modes using random actuations. 
  Spatio-temporal evolution of the KS equation and reconstruction error using 
  (a) $n^{true}_f = 64$ and 
  (b)-(e) truncated modes $n_f = \{32, 20, 16, 14\}$. 
  (f) Energy spectrum $E(\kappa_l) = \frac{1}{5000}\sum_{k = 1}^{5000}|c_l(k)|^2$, where $l = 1,2,\dots, n_f/2-1$.}
  \label{fig:mb_fo_prediction}
\end{figure}

In this section, we consider the physics-based model, i.e., the Fourier model with a truncated number of modes in Eq.~\ref{eq:spectral_ks}. 
The influence of the truncation level on the accuracy and stability of the simulation is shown in Fig.~\ref{fig:mb_fo_prediction}. 
We focus on the chaotic regime with $\nu = 0.08$ with random actuations while systematically varying $n_f$. For each case, the initial condition is generated by evolving the environment (i.e., the model with $n^{true}_f=64$) until it reaches the chaotic attractor. The state on the attractor is projected onto the lower-dimensional subspaces for $n_f=\{14,16,20,32\}$ (the solver is unstable for $n_f < 14$.). 
As the number of Fourier modes decreases, the lower-order models diverge faster from the true environment solution.

\begin{figure}
  \centering
  \includegraphics[width=\textwidth]{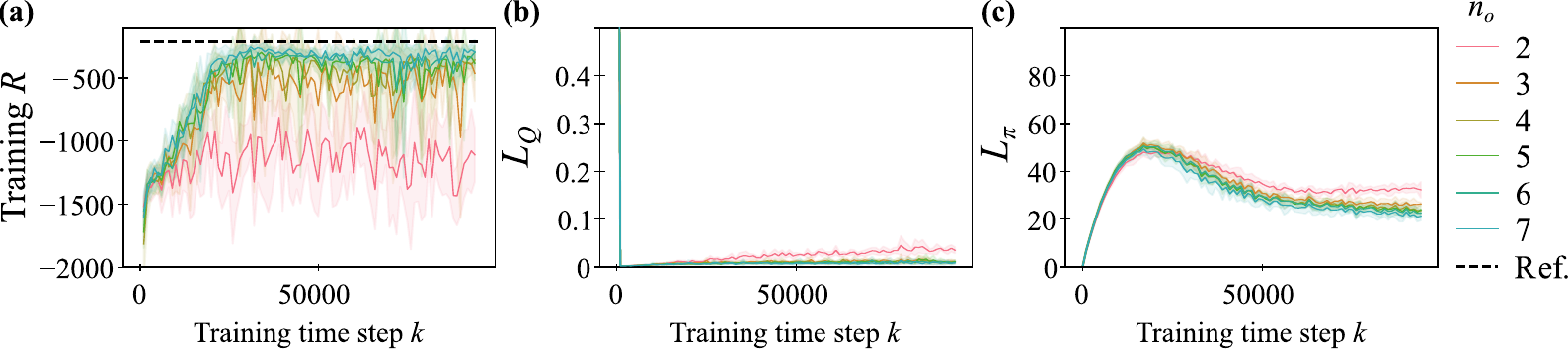}
  \vspace*{-4pt}
  \caption{Truncated Fourier model: training performance for varying number of sensors $n_o$. Mean and standard deviation over 5 runs of (a) return $R$, (b) critic loss $L_Q$, and (c) policy loss $L_\pi$. 
  Ref. indicates the maximum return achieved by the model-free algorithm with noise-free full-observability, i.e., $n_o = 64$ in Fig.~\ref{fig:mf_all}(a).}
  \label{fig:mb_fo_1}
\end{figure}

\begin{figure}
  \centering
  \includegraphics[width=.8\textwidth]{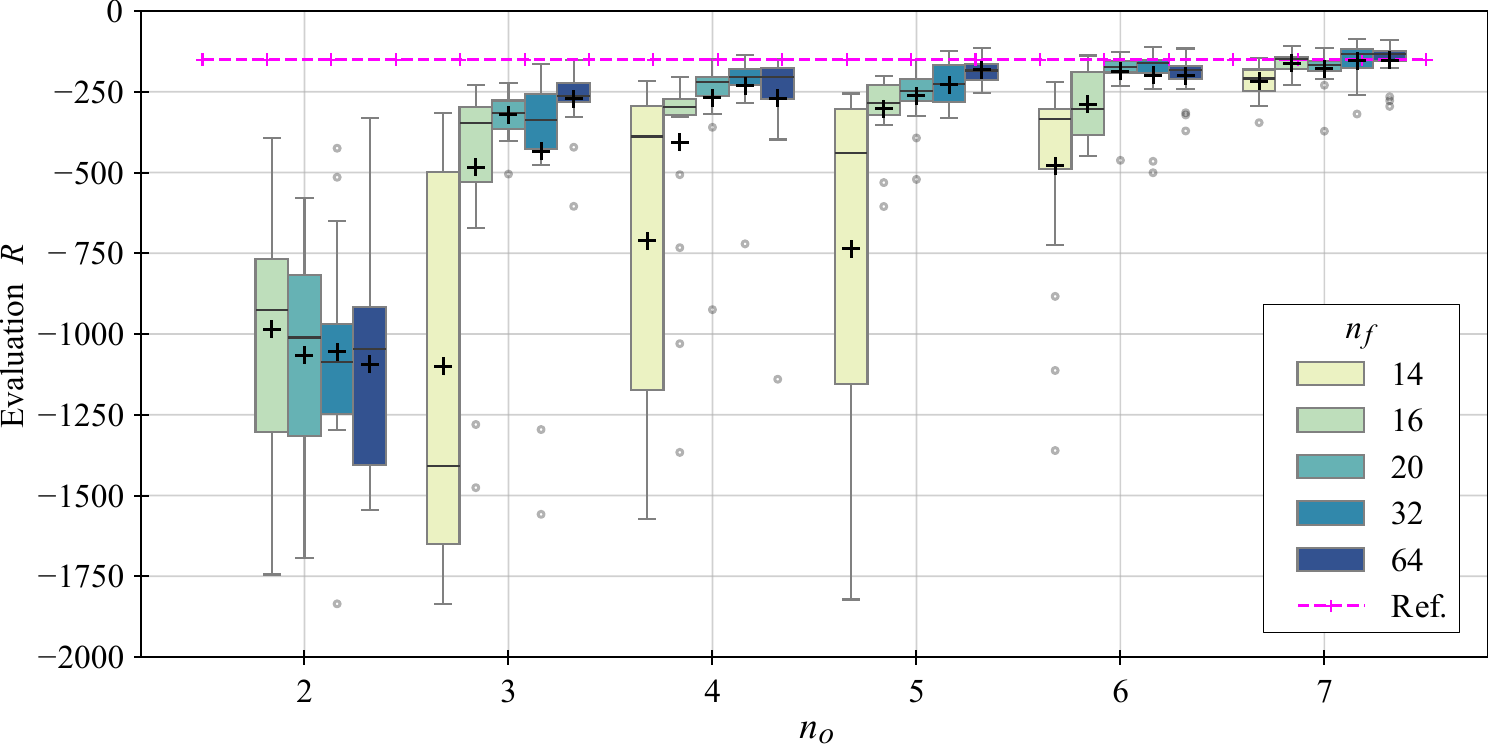}
  \vspace*{-4pt}
  \caption{
  Truncated Fourier model: 
  DA-MIRL performance 20 evaluation episodes for varying number of sensors $n_o$ and number of Fourier modes $n_f$. (The model with $n_f=14$ is numerically unstable with $n_o = 2$, hence it is not shown.)}
  \label{fig:mb_fo_eval}
\end{figure}

\begin{figure}
  \centering
  \includegraphics[width=.95\textwidth]{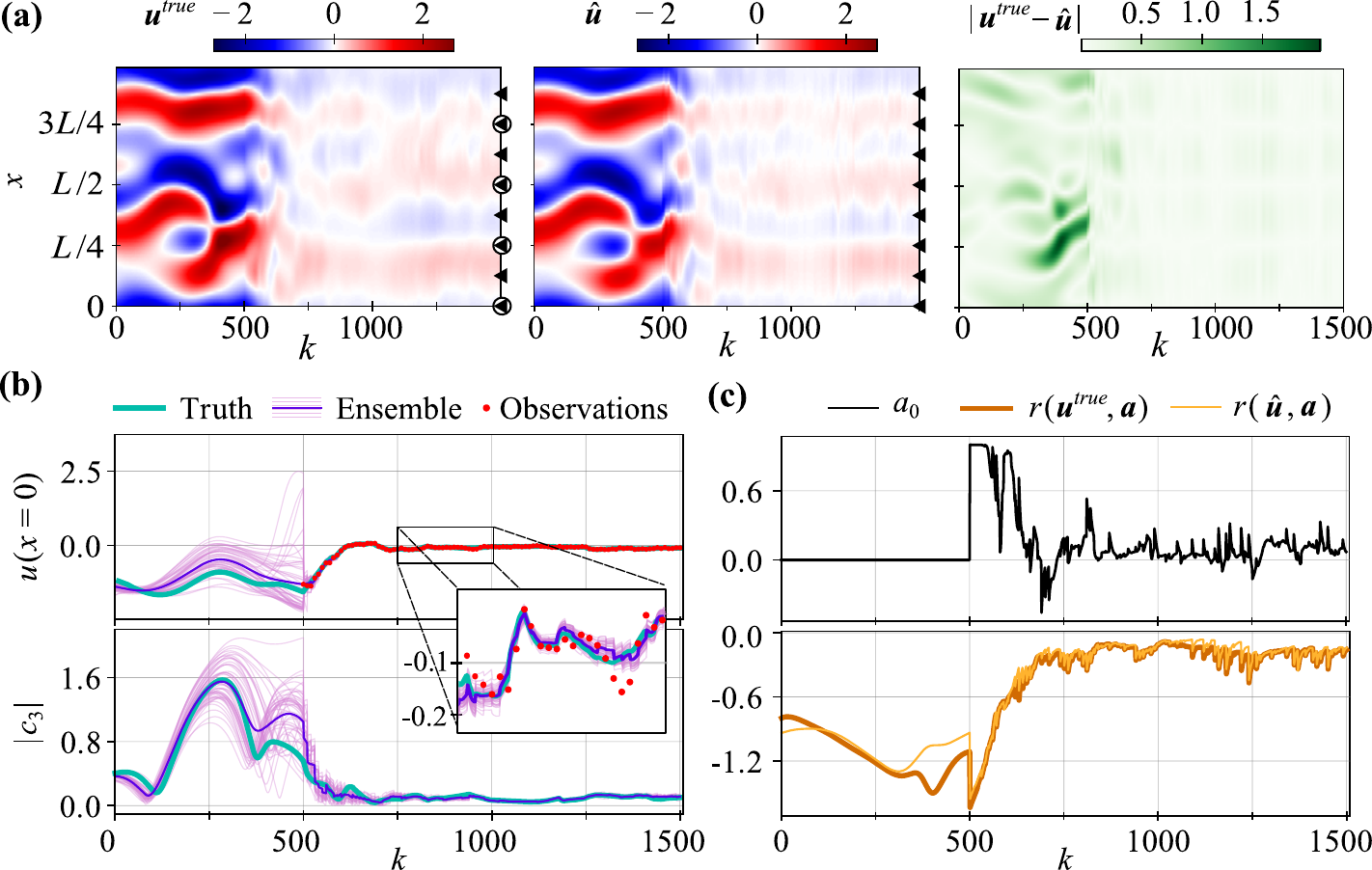}
  \vspace*{-4pt}
  \caption{
Truncated Fourier model: performance of the DA-MIRL with $n_f=16$. 
  (a) Spatio-temporal evolution of 
  the environment (the truth), 
  the model (ensemble mean), 
  and 
  the reconstruction error. The triangles and circles indicate the location of the $n_a = 8$ actuators and $n_o = 4$ sensors.
  (b) Ensemble forecast (pink), ensemble mean (purple) and true (cyan) velocity at $x=0$ and magnitude of the third Fourier coefficient. 
  (c) Agent's action at $x=0$, and 
  the reward function of the environment and the model. 
  The observations and actuations begin at $k= 500$. 
  }
\label{fig:mb_fo_controlled_ep}
\end{figure}

We set the truncation to $n_f=16$ and analyse the DA-MIRL training and control performance in Figs.~\ref{fig:mb_fo_1} and~\ref{fig:mb_fo_controlled_ep}. 
Figure~\ref{fig:mb_fo_1} shows that, in contrast to the model-free RL, the agent successfully trains with \revise{fewer} sensors than the model-free limit $n_o = 8$ (\S~\ref{sec:results_model_free}). 
Remarkably, with as few as $n_o = 3$ sensors, the proposed DA-MIRL \revise{stabilizes} and achieves episode returns that are comparable to those obtained in the idealized noise-free scenario with full observability ($n_o = 64$) in Fig.~\ref{fig:mf_all}a. 
With $n_o < 2$, there is not enough information to infer the correct state and synchronize the model with the environment, particularly as the system is driven towards the zero solution. 
Nonetheless, the actor-critic losses converge with low variance across $n_o$, compared to the model-free RL results, demonstrating the DA-MIRL robustness to partial observability. 
Next, we assess the trained agent through 20 evaluation episodes for varying number of sensors $n_o = \{2, 3, 4, 5, 6, 7\}$ and number of Fourier modes, $n_f = \{14, 16, 20, 32, 64\}$. Figure~\ref{fig:mb_fo_eval} shows the evaluation episode returns compared to the idealized baseline from the model-free RL with full state observability. 
The DA-MIRL corrects for the inaccuracy of the models with fewer Fourier modes resulting in similarly good performance across the models. 
%
Although lower-fidelity models ($n_f < 64$) intrinsically provide a limited representation of the environment, data assimilation compensates for this error, effectively bridging the gap between model predictions and environment. 
We visualize one evaluation episode with $n_o = 4$ and $n_f=16$ in Fig.~\ref{fig:mb_fo_controlled_ep}. (App.~\ref{app:random_ep} includes the results of a random episode.) The spatio-temporal plots in Fig.~\ref{fig:mb_fo_controlled_ep}a show how the model estimate of the environment remains accurate and rapidly converges to the stable solution when DA-MIRL algorithm starts at $k=500$. 
The DA-MIRL successfully recovers the true state of the environment not only at the chaotic attactor but also during the actuated transient. This is further illustrated in Fig.~\ref{fig:mb_fo_controlled_ep}b, as the ensemble spread, i.e., the uncertainty of the model, decreases with each observation. 

\subsubsection{Control-aware Echo State Network}\label{sec:results_esn}
We now focus on the data-driven modelling approach, i.e., the control-aware ESN detailed in \S~\ref{sec:model_esn}. 
The ESN is trained offline with a reservoir size of $n_h = 1000$ using full-state data. 
We generate 50 training simulations of length 1,500 time steps with different initial conditions and random actuations. 
The dataset is split into 80\% for training, 10\% for validation, and 10\% for testing. On the test set, the ESN achieves a mean relative error of approximately 3\% over short trajectories of one Lyapunov time, indicating strong short-term predictive skill. Details on the training procedure and hyperparameter selection are provided in App.~\ref{app:training}.

\begin{figure}
  \centering
  \includegraphics[width=\textwidth]{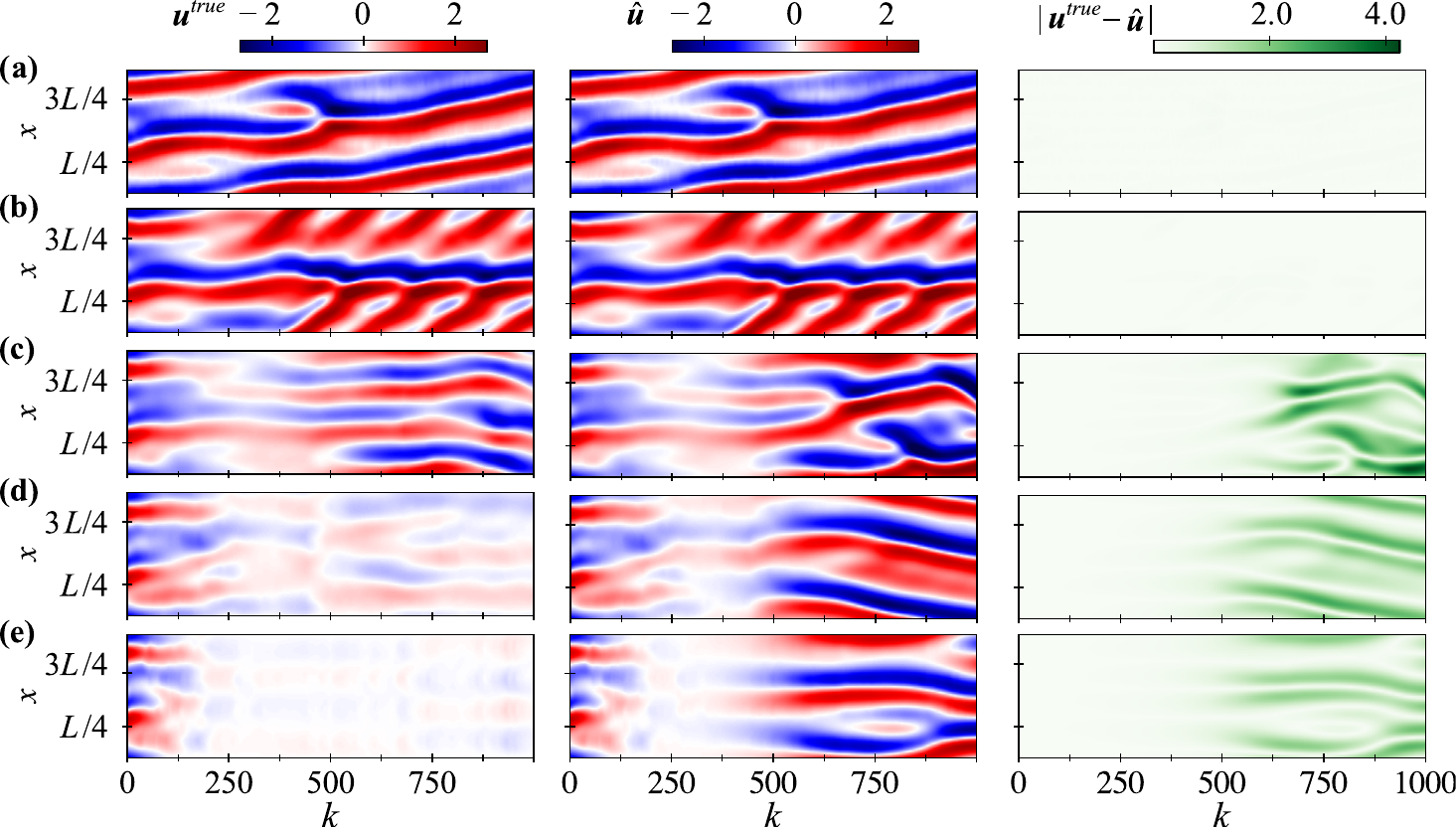}
  \vspace*{-4pt}
  \caption{Control-aware ESN prediction in closed loop (without data assimilation) throughout RL training. 
  Comparison between the environment state, the prediction of a control-aware ESN after a washout, and the mean reconstruction error. 
  The actuations are selected 
  (a) randomly (the condition in which the ESN has been trained offline), 
  (b-d) by the actor after 5, 10, 15 learning episodes where the actor is trained, and 
  (e) by the actor using the best actor weights, $\vect{\theta}^{\pi\star}$, based on the evaluation error. The actuations in (e) lead to stabilization of the flow. The predictability horizon of the ESN decreases in this unseen regime. In the online DA-MIRL training, this deviation is corrected by the data assimilation.}
  \label{fig:mb_esn_prediction}
\end{figure}

Figure~\ref{fig:mb_esn_prediction} demonstrates the ESN’s forecasting capabilities in closed-loop with the KS equation under various control scenarios. 
The reconstruction error is smallest with random actuations because the control-aware ESN is trained on this regime (Fig.~\ref{fig:mb_esn_prediction}a). 
As the actor learns to control the environment, the control-aware ESN is exposed to increasingly unfamiliar regimes, particularly as the learned policy stabilizes the flow. 
The reconstruction error remains small during the first $k=500$ for the controlled system with policies learned at different stages of the learning process (Fig.~\ref{fig:mb_esn_prediction}b-d) as well as with the optimal control policy (Fig.~\ref{fig:mb_esn_prediction}e). 
This validates the control-aware architecture proposed for the echo state network. 
The ESN’s prediction horizon decreases as the controller drives the system towards the unstable zero solution, where the dynamics have not been seen by the ESN. In our DA-MIRL framework, we overcome this limitation in the analysis step without explicitly retraining the ESN. In an experimental setting, this corresponds to training a cheap surrogate model offline based on experimental or high-fidelity simulation data, which is deployed for online \revise{reinforcement learning} with real-time observations without retraining. \revise{Integration of online learning methods for the ESN~\citep[e.g.,][]{novoa2025OnlineModelLearning} to improve its forecasting capability in unseen regimes is scope for future work.}

\begin{figure}
  \centering
  \includegraphics[width=\textwidth]{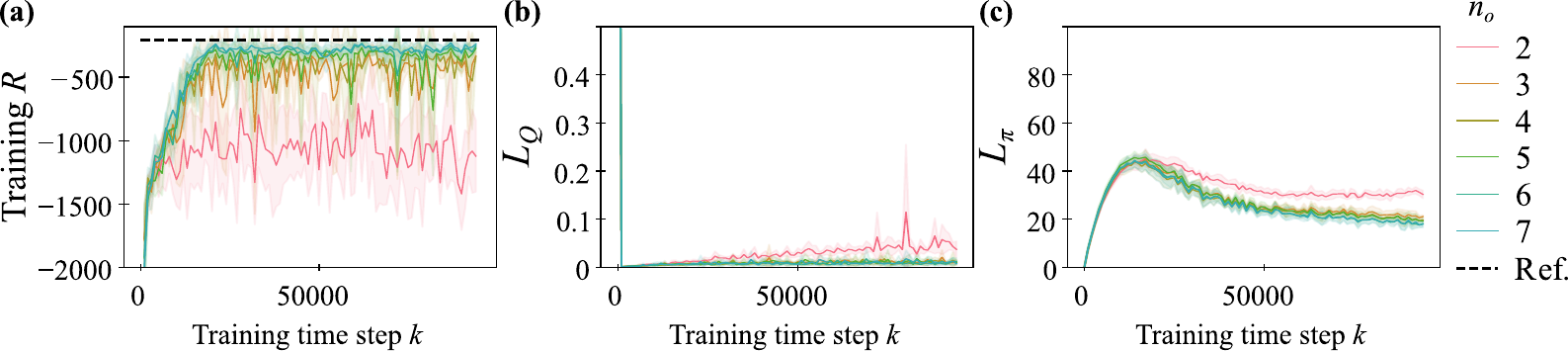}
  \vspace*{-4pt}
  \caption{Control-aware ESN: DA-MIRL training for varying number of sensors. Shown in 
  (a) non-discounted returns of training episodes, 
  (b) $Q$-loss, 
  (c) policy loss across 5 runs (mean and 1 standard deviation in lighter colour). Ref. indicates the maximum return achieved by the model-free algorithm using the full-state observations, i.e., $n_o = n_s = 64$ in~\ref{fig:mf_all}(a).
  }
  \label{fig:mb_esn_1}
\end{figure}

\begin{figure}
  \centering
  \includegraphics[width=.8\textwidth]{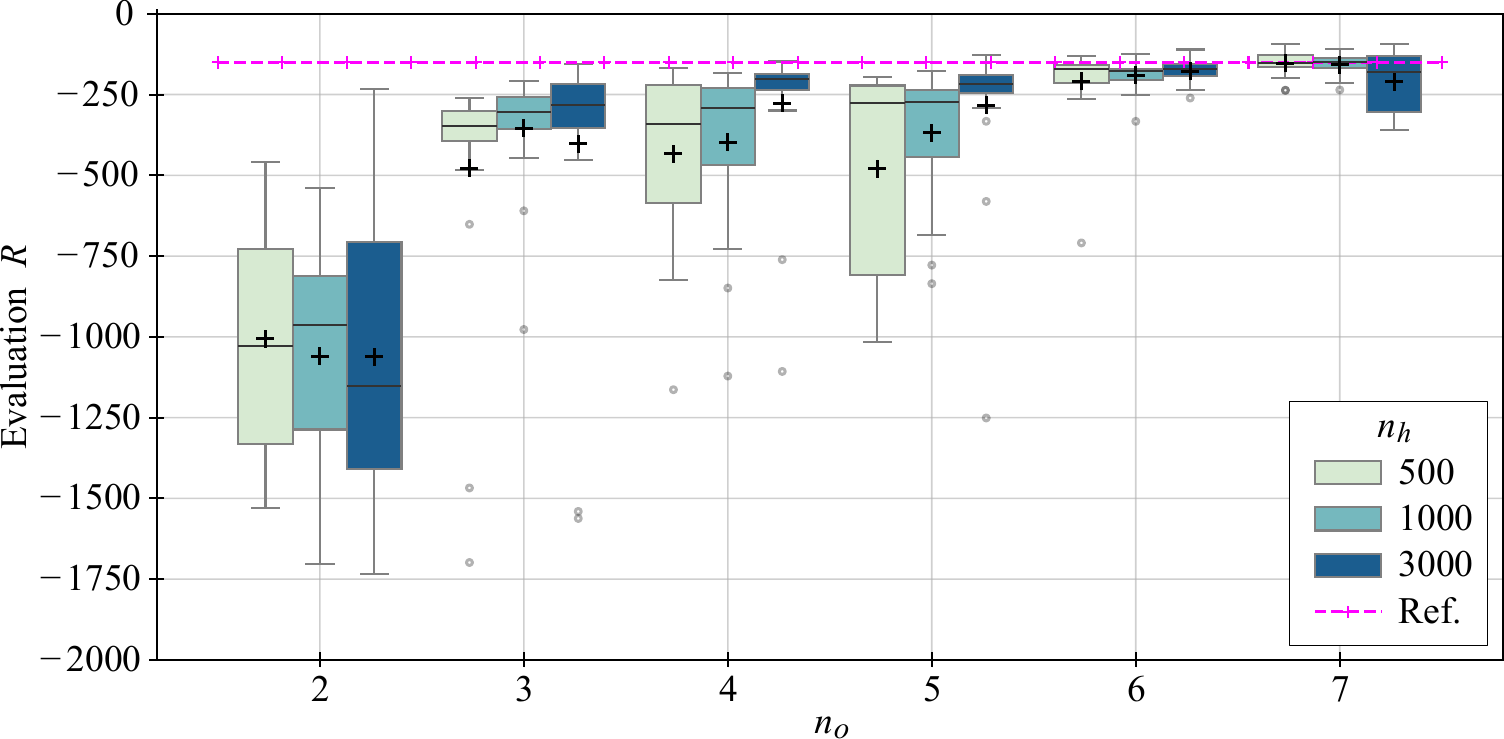}
  \vspace*{-4pt}
  \caption{Control-aware ESN: DA-MIRL performance 20 evaluation episodes 
  with varying number of sensors $n_o$ and reservoir size $n_h$. }
  \label{fig:mb_esn_eval}
\end{figure}

\begin{figure}
  \centering
  \includegraphics[width=.95\textwidth]{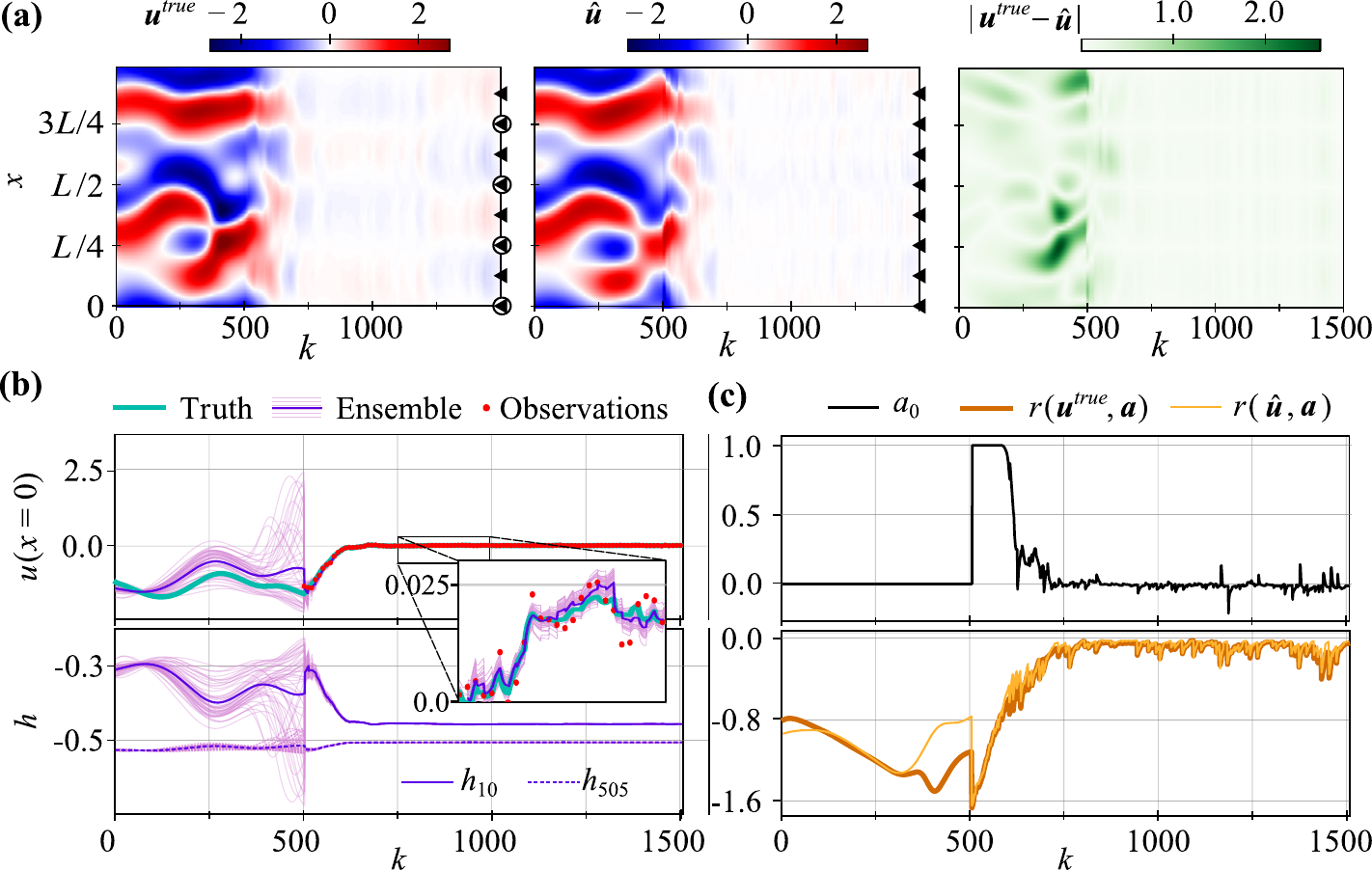}
  \vspace*{-4pt}
  \caption{Control-aware ESN: performance of DA-MIRL with $n_h = 1000$. 
  (a) Spatio-temporal evolution of 
  the environment (the truth), 
  the model (ensemble mean), 
  and 
  the reconstruction error. The triangles and circles indicate the location of the $n_a = 8$ actuators and $n_o = 4$ sensors.
  (b) Ensemble forecast (pink), ensemble mean (purple) and true (cyan) velocity at $x=0$ and 
  two reservoir states which take $u_0$ as input (no reference true value).
  (c) Agent's action at $x=0$, and 
  the reward function of the environment and the model. 
  The observations and actuations begin at $k= 500$. 
  }
  \label{fig:mb_esn_controlled_ep}
\end{figure}

Figure~\ref{fig:mb_esn_1} summarizes the training performance for the ESN-based DA-MIRL. 
Stabilization is reliably achieved with as few as $n_o = 3$ sensors, matching the efficiency observed with the truncated Fourier model in Fig.~\ref{fig:mb_fo_1}. \revise{The method is also robust to different configurations of sensors, and the results obtained with uniformly spaced sensors are representative.}
To assess the impact of sensor count and model complexity, we evaluate the trained agent over 20 evaluation episodes for $n_o = \{2, 3, 4, 5, 6, 7\}$ and reservoir sizes $n_h = \{500, 1000, 3000\}$, which shows a quick saturation of performance with increasing reservoir size. Figure~\ref{fig:mb_esn_eval} shows the resulting episode return statistics compared with the idealized model-free agent. 
Figure~\ref{fig:mb_esn_controlled_ep} shows an evaluation episode with $n_o = 4$. (App.~\ref{app:random_ep} includes the results of a random episode.) 
The control-aware ESN predicts the spatio-temporal evolution of the controlled KS equation and yields similar reconstruction error to the physics-based truncated Fourier model, which shows the efficacy of the proposed DA-MIRL. 
Figure~\ref{fig:mb_esn_controlled_ep}(b) highlights the evolution of the true environment state and the model ensemble prediction for one of the four observables, i.e., $u(x=0, t)$. The EnKF rapidly leads the ensemble to the true solution as observations become available at $k=500$. \revise{While the uncertainty of the estimate is higher at the unobserved locations, the reconstruction converges to the ground truth across the spatial domain (Fig.~\ref{fig:mb_esn_controlled_ep}a).}
\revise{Fig.~\ref{fig:mb_esn_controlled_ep}(b)} also \revise{illustrates} how the ESN’s internal dynamics echo the observed environment: the evolution of two reservoir state ensembles, $h_{10}$ and $h_{505}$, which take $u(x=0)$ as input, are a transformation of the physical state $u(x=0)$. 
The close overlap between the true reward and the model estimated reward (Fig.~\ref{fig:mb_esn_controlled_ep}c), quantitatively validates the proposed DA-MIRL framework's ability to maintain policy performance despite partial observability. \revise{Ultimately, for this initial condition and experiment setup, the data-driven model with $n_h = 1000$ yields a higher episode return ($R = -226$) than the coarse-grained model with $n_f = 16$ ($R = -296$) in Fig.~\ref{fig:mb_fo_controlled_ep}, with a smaller error from the target of zero state. However, their evaluation across different initial conditions (Figs.~\ref{fig:mb_fo_eval} and ~\ref{fig:mb_esn_eval}) indicate that their performances are comparable on average, where increasing the number of observations compensates for lack of model fidelity.}

\begin{figure}
  \centering
  \includegraphics[width=\textwidth]{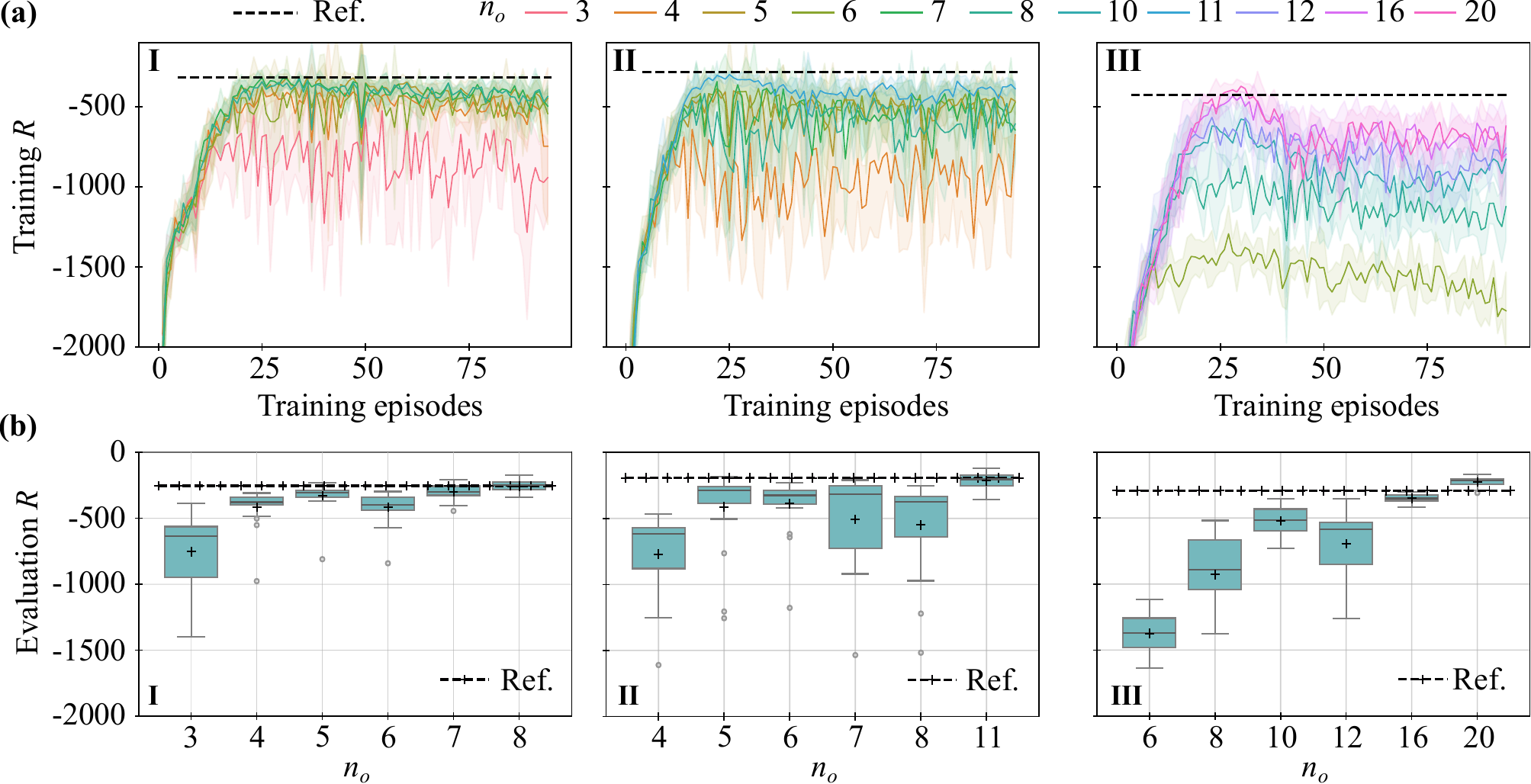}
  \vspace*{-4pt}
  \caption{
  Control-aware ESN: DA-MIRL performance over increasingly chaotic regimes with viscosity (I) $\nu = 0.05$, (II) $\nu = 0.03$, (II) $\nu = 0.01$. 
  Episode return
  (a) evolution during training, and 
  (b) statistics over 20 evaluation episodes.}
  \label{fig:mb_esn_nu}
\end{figure}

Lastly, 
to assess robustness, we perform the analysis on more chaotic regimes, which are obtained by decreasing the viscosity parameter $\nu = \{0.05, 0.03, 0.01\}$ ($L \approx \{28, 36, 63\}$, respectively). 
Without actuation, the system with $\nu = 0.05$ converges to a fixed point after a chaotic transient, in contrast, with $\nu = \{ 0.03, 0.01\}$ the system is unstable and increasingly chaotic with Lyapunov times $LT \approx \{250, 235\}$ time steps, respectively. 
The control-aware ESN (with fixed $n_h = 1000$) is retrained for each regime and continues to capture the essential dynamics. In contrast, the truncated Fourier model requires more modes for stability as chaos intensifies. Figure~\ref{fig:mb_esn_nu} presents the training and evaluation episode returns across these regimes.
Before starting the observations and actuations in the DA-MIRL, we forecast the ensemble in closed loop for $LT$ units (350 time steps for the $\nu = 0.05$ case). 
As the KS system has increasingly chaotic behaviour and the dominant wavenumber increases, the state estimation and control tasks become more challenging. 
On the one hand, the minimum number of actuators $n_a$ required for stabilization increases with the system’s complexity. We find that the required actuators for chaos synchronization are $n_a = \{9, 12, 21\}$ for $\nu=\{0.05, 0.03, 0.01\}$ respectively. (Exploring further the controllability as a function of the number or location of the actuators is left for future work.)
On the other hand, the DA-MIRL requires more sensor data for accurate reconstruction in highly chaotic scenarios. As expected, state estimation becomes challenging in the most chaotic regime ($\nu = 0.01$). To mitigate the challenges of increased level of chaos in the system, we increase the ensemble size and the inflation factor ($m = 200$ and $\rho_I = 1.05$). 
We find that the DA-MIRL framework generalizes and is robust to partial observability and noise across different dynamics.

\section{Conclusion}\label{sec:conclusion}

Controlling spatio-temporally chaotic systems is a challenging task because of nonlinearity, high-dimensionality, and unpredictability. 
In experiments and practical applications, data are collected from a limited number of sensors, which provide noisy and partial observations. 
Although reinforcement learning (RL) has been successfully applied for the control of chaotic systems, it may not be robust under noisy and partially observed systems. 
%
%
We propose the data-assimilated model-informed reinforcement learning (DA-MIRL) framework, which transforms the partially observable Markov decision process into a fully observed process by  integrating
(i) a predictive model of the system's dynamics, which effectively replaces the environment in the RL loop; 
(ii) sequential data assimilation for real-time state estimation from observations, here, the ensemble Kalman Filter (EnKF); and 
(iii) an off-policy actor-critic RL algorithm to learn control policies, here, the Deep Deterministic Policy Gradient (DDPG). 
We test the DA-MIRL on the Kuramoto-Sivashinsky (KS) equation, which is a partial differential equation with spatio-temporally chaotic solutions. 
The control strategy aims to achieve two primary objectives: stabilizing the system's chaotic dynamics toward a zero-energy equilibrium whilst minimizing the actuation power required for stabilization. 
%
%
We test the framework with two different models for the environment:  
(i) a physics-based low-order model (a coarse-grained discretization of the equation); and 
(ii) a data-driven approach with echo state networks (ESNs). 
We propose a control-aware formulation of the ESN that enables the prediction of controlled dynamical systems. Although the control-aware ESN is trained on the KS system with random actuation, the network generalizes to forecast the dynamics under different control strategies and regimes. 
The DA-MIRL is robust to noisy partial observations, for which model-free RL agents may fail. The DA-MIRL learns control policies with 60\% fewer sensors than required by model-free RL. 
The results are consistent for both the physics-based and the data-driven models of the environment. 
Finally, we show that the robustness to noise and partial observability across a range of chaotic regimes.    
This work opens up possibilities for RL-based control of chaotic systems from partial and noisy observations.

\paragraph{Acknowledgements}
We acknowledge the support from 
the UKRI AI for Net Zero grant EP/Y005619/1,
the ERC Starting Grant PhyCo 949388, 
and the grant EU-PNRR YoungResearcher TWIN ERC-PI\_0000005. The authors would like to thank Max Weissenbacher for valuable insights and for the numerical implementation of the Kuramoto-Sivashinsky equation.

\paragraph{Code availability}
The code is available as a GitHub repository at \href{https://github.com/MagriLab/DA-RL}{MagriLab/DA-RL}. The individual runs can also be viewed at Weights \& Biases for model-free RL (\href{https://wandb.ai/defneozan/DA-RL-MF}{DA-RL-MF}), DA-MIRL with the truncated Fourier model (\href{https://wandb.ai/defneozan/DA-RL-MB-Fo}{DA-MIRL-Fo}), and with the ESN (\href{https://wandb.ai/defneozan/DA-RL-MB-ESN}{DA-MIRL-ESN}). 

\appendix
\section{A derivation of the EnKF}\label{app:EnKF}

Assuming Gaussian likelihood and prior terms in \eqref{eq:Bayes}, i.e., 
\begin{subequations}\nonumber
\begin{align}
  \mathcal{P}(\vect{\psi}) &\propto \exp\left\{-\frac{1}{2}\left(\vect{\psi}^\mathrm{f} - \vect{\psi}\right)^\top \matr{C}_{\psi\psi}^{\mathrm{f}^{-1}} \left(\vect{\psi}^\mathrm{f} - \vect{\psi}\right)\right\}, \;\text{and}\\
    \mathcal{P}(\vect{o}\mid \vect{\psi}) &\propto \exp\left\{-\frac{1}{2}\big(\vect{o} - \hat{\matr{M}}\vect{\psi}\big)^\top \matr{C}_{oo}^{-1} \big(\vect{o} - \hat{\matr{M}}\vect{\psi}\big)\right\},
\end{align}
\end{subequations}
where $\hat{\matr{M}} = [\matr{0}|\matr{I}]$ such that $\hat{\matr{M}}\vect{\psi}=\matr{M}(\vect{s})$, then the posterior is $\mathcal{P}(\vect{\psi}\mid \vect{o}) \propto 
  \exp{\left\{-\mathcal{J}(\vect{\psi})\right\}}$, where
\begin{align}
  \mathcal{J}(\vect{\psi}) = \left(\vect{\psi}^\mathrm{f} - \vect{\psi}\right)^\top \matr{C}_{\psi\psi}^{\mathrm{f}^{-1}} \left(\vect{\psi}^\mathrm{f} - \vect{\psi}\right)
  +\big(\vect{o} - \hat{\matr{M}}\vect{\psi}\big)^\top \matr{C}_{\epsilon\epsilon}^{-1} \big(\vect{o} - \hat{\matr{M}}\vect{\psi}\big)
\end{align}
In a MAP approach, the most likely state $
  \vect{\psi}^\mathrm{a}=\argmax_{\vect{\psi}}{\mathcal{P}(\vect{\psi}\mid \vect{o})} 
$, or equivalently, because of the Gaussian assumption, $\vect{\psi}^\mathrm{a}= \argmax_{\vect{\psi}}\log{\mathcal{P}(\vect{\psi}\mid \vect{o})} = \argmin_{\vect{\psi}}{\mathcal{J}}$. 
Therefore,
\begin{align}
\dfrac{\mathrm{d}~\mathcal{J}}{\mathrm{d}~\vect{\psi}_j}\bigg|_{\vect{\psi}_j^\mathrm{a}}=
2\,\matr{C}^{\mathrm{f}^{-1}}_{\psi\psi}\left(\vect{\psi}_j^\mathrm{a}-\vect{\psi}_j^\mathrm{f}\right)-2\,\hat{\matr{M}}^\top\matr{C}^{-1}_{oo}\left(\vect{o}_j-\hat{\matr{M}}\vect{\psi}_j^\mathrm{a}\right)=0\nonumber\\
\therefore \left(\matr{C}^{\mathrm{f}^{-1}}_{\psi\psi} + \hat{\matr{M}}^\top\matr{C}^{-1}_{oo}\hat{\matr{M}}\right)\vect{\psi}_j^\mathrm{a}=\matr{C}^{\mathrm{f}^{-1}}_{\psi\psi}\vect{\psi}_j^\mathrm{f}+\hat{\matr{M}}^\top\matr{C}^{-1}_{oo}\vect{o}_j. 
\end{align}
Making use of the Woodbury formula, this expression simplifies to 
\begin{align}
\vect{\psi}_j^\mathrm{a} 
&= \vect{\psi}_j^\mathrm{f}+\left(\matr{C}_{\psi\psi}^\mathrm{f}\hat{\matr{M}}^\top\right)\left(\matr{C}_{dd}+\hat{\matr{M}}\matr{C}_{\psi\psi}^\mathrm{f}\hat{\matr{M}}^\top\right)^{-1}\left(\vect{o}_j-\hat{\matr{M}}\vect{\psi}_j^\mathrm{f}\right). 
\end{align}
Evaluating the terms with $\hat{\matr{M}}$, we obtain the EnKF equation~\ref{eq:EnKF}. 

\section{Training details}\label{app:training}

Here, we discuss further details of the RL agent and control-aware ESN training procedures. 
Tab.~\ref{tab:training_params} includes a summary of training parameters of DA-MIRL for the $\nu = 0.08$ case, the set of optimal hyperparameters for the ESN with reservoir size of $n_h = 1000$, and the average computation times for different experiments. 
The training was performed on an NVIDIA Quadro RTX 8000 GPU. 
We leverage the parallelization and Just-In-Time (JIT) compilation capabilities of JAX for computationally-efficient ensemble forecasting and ESN training.

\subsection{DDPG agent}
\begin{algorithm} 
  \caption{Update DDPG networks.}\label{alg:ddpg}
    \begin{algorithmic}[1]
      \State \textbf{Sample batch} of $n_{b}$ random tuples $\left(\vect{s}_k^{(i)}, \vect{a}_k^{(i)}, r_k^{(i)}, \vect{s}_{k+1}^{(i)}\right), \; i = 1, 2, \dots, n_b $ from $\mathcal{B}$
      \State \textbf{Set temporal difference error for critic}: 
      \begin{align*}
       \delta^{(i)} = r_k^{(i)} + \gamma Q^{\pi}{'}\left(\vect{s}_{k+1}^{(i)}, \vect{\pi}'\left(\vect{s}_{k+1}^{(i)};\vect{\theta}^{\pi'}\right);\vect{\theta}^{Q'}\right) - Q^\pi\left(\vect{s}_k^{(i)}, \vect{a}_k^{(i)};\vect{\theta}^{Q}\right) 
      \end{align*}
      \State \textbf{Update critic} by minimizing loss: 
      \begin{align*}
        L_Q = \frac{1}{n_b} \sum_{i = 1}^{n_b} \left(\delta^{(i)}\right)^2
      \end{align*}
      Compute the gradient $\nabla_{\vect{\theta}^Q} L_Q$ by backpropagation and apply gradient descent on $\vect{\theta}^Q$.
      \State \textbf{Update actor} by minimizing the loss: 
      \begin{align*}
        L_\pi = -\frac{1}{n_b} \sum_{i = 1}^{n_b} Q^\pi\left(\vect{s}_k^{(i)}, \vect{\pi}\left(\vect{s}_k^{(i)};\vect{\theta}^\pi\right);\vect{\theta}^Q\right).
      \end{align*}
      Compute the policy gradient $\nabla_{\vect{\theta}^\pi} L_\pi$ using backpropagation and the chain rule: 
      \begin{align*}
        \nabla_{\vect{\theta}^\pi} Q^\pi\left(\vect{s}, \vect{\pi}\left(\vect{s};\vect{\theta}^\pi\right);\vect{\theta}^Q\right) = \nabla_{\vect{a}} Q^\pi\left(\vect{s}, \vect{a};\left.\vect{\theta}^Q\right)\right\rvert_{\vect{a} = \vect{\pi}(\vect{s};\vect{\theta}^\pi)} \nabla_{\vect{\theta}^\pi} \vect{\pi}\left(\vect{s};\vect{\theta}^\pi\right)
      \end{align*}
      Apply gradient descent on $\vect{\theta}^\pi$.
      \State \textbf{Update target networks}: 
      \begin{align*}
      \vect{\theta}^{Q'} &\leftarrow \tau \vect{\theta}^Q + (1 - \tau)\vect{\theta}^{Q'}\\
      \vect{\theta}^{\pi'} &\leftarrow \tau \vect{\theta}^\pi + (1 - \tau)\vect{\theta}^{\pi'}
      \end{align*}
    \end{algorithmic}
  \vspace*{-4pt} 
\end{algorithm}
\begin{algorithm}[!htb]
  \caption{Training procedure}\label{alg:training}
  \begin{algorithmic}[1]
    \State \textbf{Stage 1: Random Actuation} \hfill \Comment{Replay buffer warm-up}
    \State \quad Apply random actuation $\vect{a}_k \sim \mathcal{U}(-1,1)$ and store $(\vect{s}_k, \vect{a}_k, r_k, \vect{s}_{k+1})$ in $\mathcal{B}$
    \For{episode = 1 to training episodes}
      \State \textbf{Stage 2: Learning} \hfill \Comment{Policy exploration}
      \State \quad Apply control policy $\vect{a}_k = \vect{\pi}(\vect{s}_k;\vect{\theta}^\pi) + \mathcal{N}_k$
      and store $(\vect{s}_k, \vect{a}_k, r_k, \vect{s}_{k+1})$ in $\mathcal{B}$
      \State \quad Update actor and critic networks every time step using Alg.~\ref{alg:ddpg}
      \If {mod(episode, evaluation frequency) = 0}
      \State \textbf{Stage 3: Evaluation} \hfill \Comment{Policy evaluation}
      \State \quad Apply control policy $\vect{a}_k = \vect{\pi}'(\vect{s}_k;\vect{\theta}^{\pi'})$
      \If {$R_{eval} > R_{eval}^\star$} $R_{eval}^\star \leftarrow R_{eval}$, $\vect{\theta}^{Q\star} \leftarrow \vect{\theta}^{Q'}$, $\vect{\theta}^{\pi\star} \leftarrow \vect{\theta}^{\pi'}$
      \EndIf
      \EndIf
    \EndFor
  \end{algorithmic}
  \vspace*{-4pt}
\end{algorithm}
  
\revise{The implementation of the DDPG algorithm follows~\citet{lillicrap2016ContinuousControlDeep}. The update of the DDPG networks using a random batch of $(\vect{s}_k, \vect{a}_k, r_k, \vect{s}_{k+1}) 
\in \mathcal{B}$ of batch size $n_b$ is given in~Alg.~\ref{alg:ddpg}.} The agent in the DA-MIRL framework is trained episodically in three stages, as summarized in Alg.~\ref{alg:training}. The episodes in the stages are referred to as (i) randomly actuated, (ii) exploration-learning, and (iii) evaluation episodes. 
Random re-initialization of the episodes enable the agent to gain experience in different portions of the state space. For each episode, the ensemble is initialized around the attractor, and then propagated for one Lyapunov time to allow for sufficient divergence of the trajectories.
Prior to training, we start filling the replay buffer with samples from episodes generated by applying random actuations. We then initiate the learning with network updates, while applying the control policy given by the actor network perturbed by Gaussian noise to promote exploration. 
At intervals, we run evaluation episodes using the control policy given by the target actor network. We save the weights that results in the best evaluation return of the model. In Alg.~\ref{alg:da_mbrl}, we show details of a learning episode. The procedure is identical for the other stages except for how the actions are determined.
\subsection{Control-aware ESN}
Training the ESN consists of finding the output weights $\matr{W}_{out}$. To this aim, we initialize the reservoir as $\vect{h}(t_0) = \vect{0}$, and
compute the reservoir states 
$\matr{H} = [\vect{h}(t_1) \; \vect{h}(t_2) \; \dots \; \vect{h}(t_{N_{t}})] \in \mathbb{R}^{n_h \times N_{t}}$ in open loop according to~\eqref{eq:control_esn_step} with the training data as input. 
We discard an initial time window (the washout) of length $k_{wash}$ to eliminate the effects of the initialization of the reservoir state and to synchronize the reservoir with the input state. Next, we solve for $\matr{W}_{out}$~\citep{lukosevicius2012PracticalGuideApplyinga}
\revise{\begin{equation}
  \matr{W}_{out}^\star = \underset{\matr{W}_{out}}{\mathrm{arg \, min}}\; \frac{1}{n_u}\sum_{i = 1}^{n_u}\left(\sum_{k = 1}^{N_{t}}(u_i(t_k)-\hat{u}_i(t_k))^2 +\lambda||\vect{w}_{out, i}||^2\right),
\label{eq:wout_optim}
\end{equation}}
where $\hat{\vect{u}}(t_k)$ is given by~\eqref{eq:control_esn_readout}, $\vect{w}_{out, i}$ is the $i^{th}$ row of $\matr{W}_{out}$, and $\lambda$ is the Tikhonov coefficient, and the $\ell_2$ regularization prevents overfitting. The solution is computed in closed form ~\citep{lukosevicius2012PracticalGuideApplyinga}
\revise{\begin{equation}
  \matr{W}_{out}^\star = \matr{U}\tilde{\matr{H}}^\top(\tilde{\matr{H}}\tilde{\matr{H}}^\top+\lambda\matr{I})^{-1},
\label{eq:wout_soln}
\end{equation}}
where $\revise{\matr{U}} = \left[\vect{u}(t_1) \; \vect{u}(t_2) \;\dots \; \vect{u}(t_{N_{t}})\right], \; \matr{U} \in \mathbb{R}^{n_u \times N_{t}}$, $\tilde{\matr{H}} = [\matr{H}; \vect{1}^\top], \; \tilde{\matr{H}} \in \mathbb{R}^{(n_h+1) \times N_{t}}$, $\vect{1}$ is a vector of ones, and $\matr{I}$ is the identity matrix.
The hyperparameters $\{\alpha, \xi_{y},
\xi_{a}, \rho, \lambda \}$ are selected through a validation scheme utilizing Bayesian optimization~\citep{racca2021RobustOptimizationValidation}. 
To this aim, we evaluate the closed-loop performance on a validation set with 5 unseen episodes. We repeat this for 3 realizations of $\matr{W}$ and $\matr{W}_{in}$, and 3 folds, i.e., trajectories, of length $N_{val}=500$ ($\sim 1$ Lyapunov time) starting from random time steps in the dataset. The validation error to minimize is 
\revise{\begin{equation}\label{eq:rel_l2}
  e_{val} = \sqrt{\frac{\sum_{k=1}^{N_{val}}\sum_{i=1}^{n_u}(u_i(t_k)-\hat{u}_i(t_k))^2}{\sum_{k=1}^{N_{val}}\sum_{i=1}^{n_u}u_i^2(t_k)}},
\end{equation}}
which we average over the number of folds, validation episodes, and realizations.
\begin{table}[h!]
  \caption{Training DA-MIRL parameters and optimal control-aware ESN hyperparameters. 
  }\label{tab:training_params}
  \centering
   \begin{tabular}{|c|c|}
   \hline
  \textbf{DA-MIRL Agent - DDPG} & \textbf{Control-aware ESN}\\
  \hline
   \begin{tabular}{ll}
    Episode length $k_{ep}$ & 1000 \\
    Control-free time steps ($k_{start}$)& 500\\ 
    Random actuation episodes & $5$\\
    Training episodes & $95$ \\ 
    Buffer capacity & $10^5$ \\ 
    Actor hidden units & [256, 256] \\ 
    Critic hidden units & [256, 256] \\ 
    Activation function & ReLU \\ 
    Actor learning rate & $3\times10^{-4}$ \\ 
    Critic learning rate & $3\times10^{-4}$ \\ 
    Discount factor $\gamma$ & 0.99 \\ 
    Soft update rate $\tau$ & 0.005 \\ 
    Batch size $n_b$ & 256 \\ 
    Exploration covariance $\matr{C}_{aa}$ & $0.2^2\matr{I}$
  \end{tabular}
  & 
  \begin{tabular}{ll}
  Washout time steps & 100 \\
  Reservoir size $n_h$ & 1000 \\
  Connectivity $n_{conn}$ & 3 \\
  Training time steps & $56 \times 10^3$ \\
  \\
  \multicolumn{2}{c}{\textit{Optimized hyperparameters}}\\
  Leak rate $\alpha$ & 0.23 \\
  Tikhonov coefficient $\lambda$ & $10^{-6}$ \\
  Spectral radius $\rho$ & 0.07 \\
  Input state scaling $\xi_{y}$ & 0.23 \\
  Input action scaling $\xi_{a}$ & 0.51 \\ \\ \\\\
  \end{tabular}
  \\
  \hline
  \multicolumn{2}{|c|}{\textbf{Training times}} \\
  \hline
  \multicolumn{2}{|c|}{
    \begin{tabular}{ll}
      Model-free RL with full state observation $n_o = 64$ & 7 min. \\
      DA-MIRL with truncated Fourier model $n_f = 16$ & 7 min. \\
      Validation of control-aware ESN $n_h = 1000$ & 4 min. \\
      Training of control-aware ESN $n_h = 1000$ & 5 s. \\
      DA-MIRL with control-aware ESN $n_h = 1000$ & 14 min. \\
    \end{tabular}
  } \\
  \hline
  \end{tabular}
\vspace*{-4pt} 
\end{table}

\section{State estimation performance with random actuations}\label{app:random_ep}
\begin{figure}
  \centering
  \includegraphics[width=.95\textwidth]{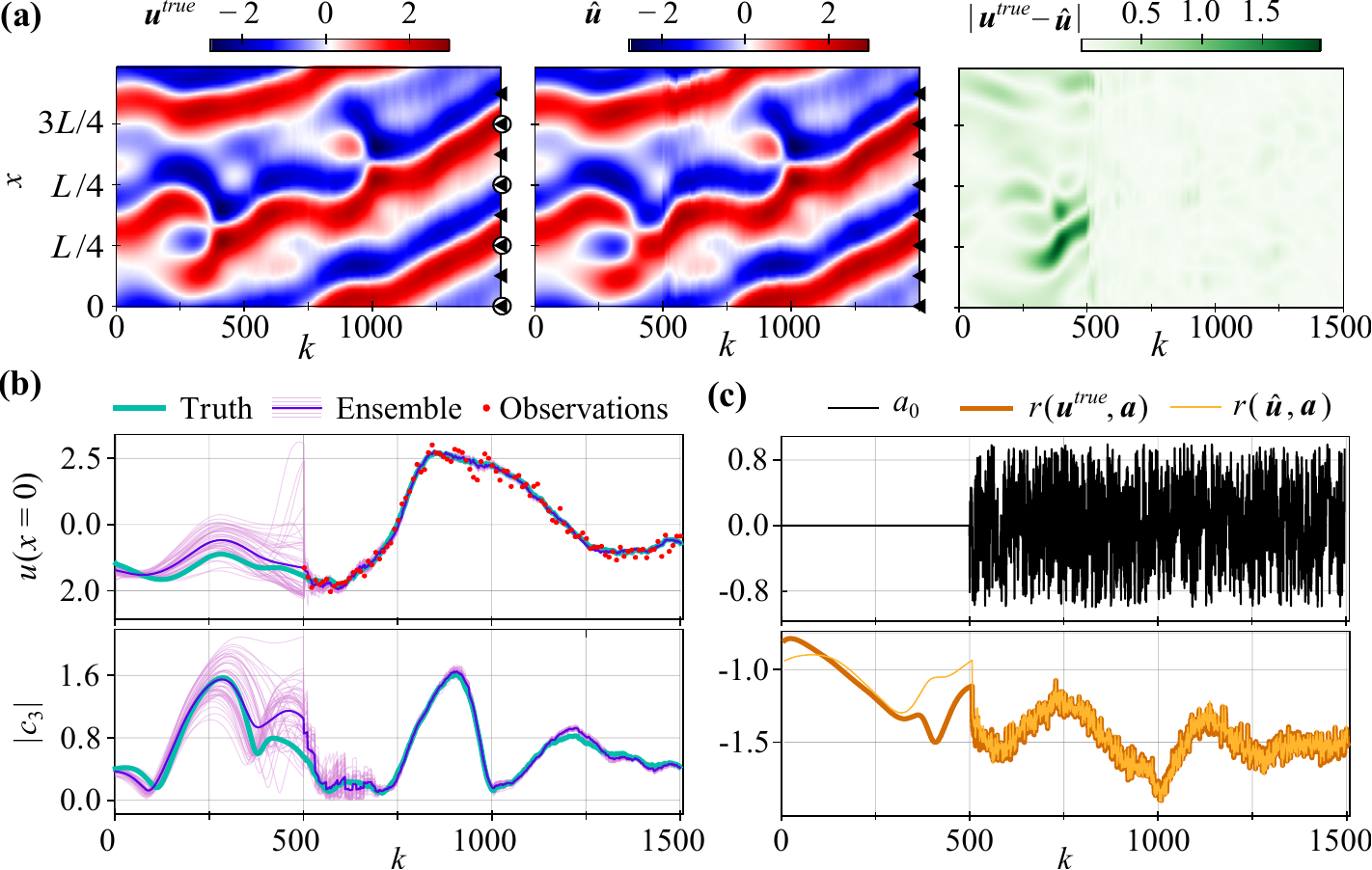}
  \vspace*{-4pt}
  \caption{
  State estimation on the truncated Fourier model with $n_f=16$ with random actuations. 
   (a)-(c) same as Fig.~\ref{fig:mb_fo_controlled_ep}. 
  }
  \label{fig:mb_fo_random_ep}
\end{figure}

\begin{figure}
  \centering \includegraphics[width=.95\textwidth]{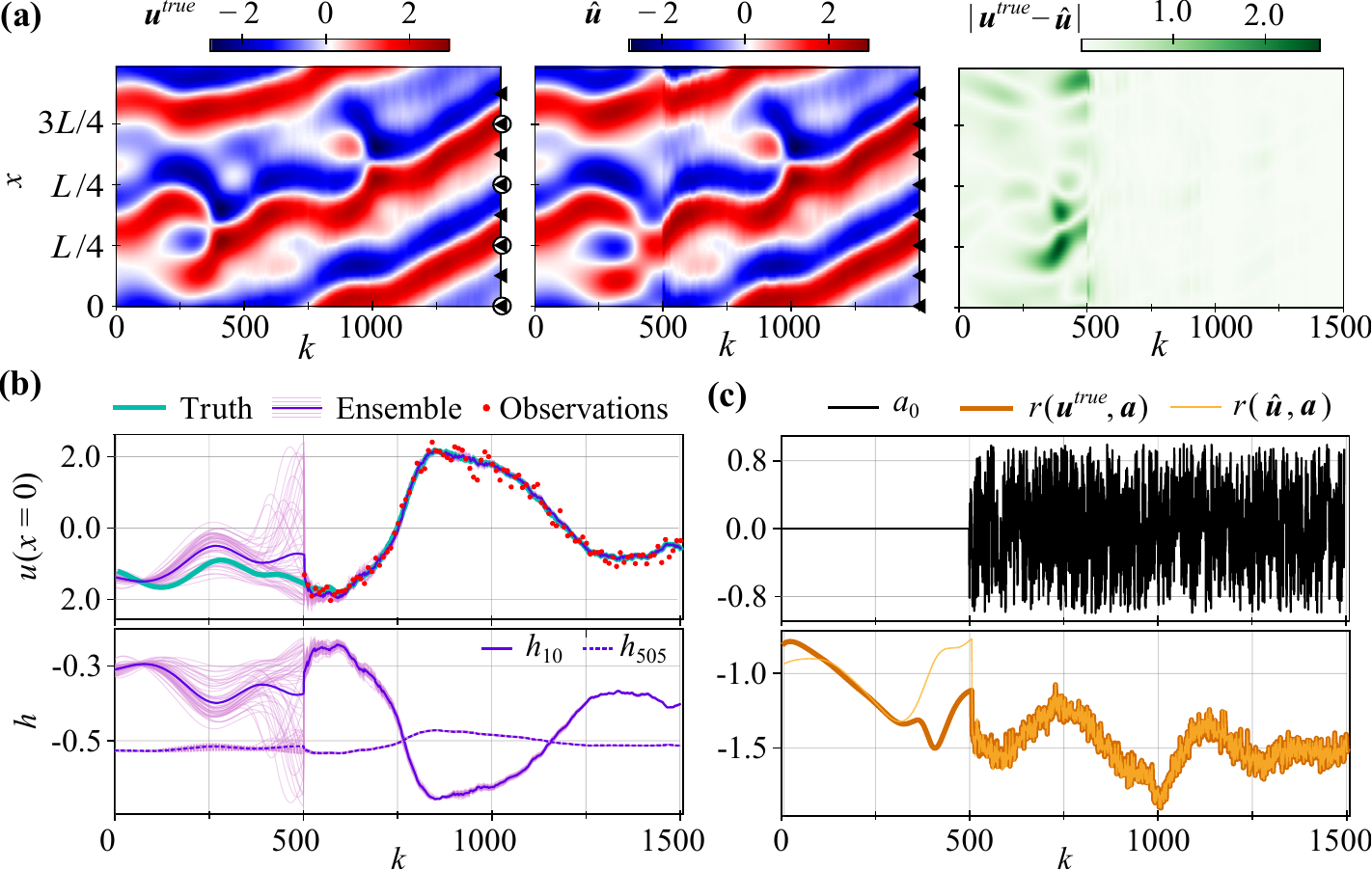}
  \vspace*{-4pt}
  \caption{
  State estimation on the control-aware ESN with $n_h = 1000$ with random actuations. (a)-(c) same as Fig.~\ref{fig:mb_esn_controlled_ep}. 
  }
  \label{fig:mb_esn_random_ep}
\end{figure}

The performance of the state estimation on random episodes, i.e., episodes in which the system is actuated with random actions is shown in Fig.~\ref{fig:mb_fo_random_ep} for the truncated Fourier model, and Fig.~\ref{fig:mb_esn_random_ep} for the control-aware ESN. 

\bibliography{bibliography} 

\end{document}